%% file: main.tex
\documentclass[trackchanges, twocolumn]{aastex7}

\usepackage{amsmath}
\usepackage{tikz}
\usepackage{subcaption}
\usetikzlibrary{arrows.meta}
\usepackage{rotating}
\usepackage[acronym]{glossaries}
\glsdisablehyper
\makeglossaries
\usepackage{fontawesome5} 
\usepackage{enumitem}

\newcommand{\mrm}[1]{\mathrm{#1}}
\newcommand{\mbf}[1]{\mathbf{#1}}
\newcommand{\mcal}[1]{\mathcal{#1}}

\newcommand{\mbs}[1]{\boldsymbol{#1}}

\NewDocumentCommand{\package}{m o}{%
  \IfNoValueTF{#2}{}{\href{#2}{\faCode}}~%
  \texttt{#1}\;%
}
\newcommand{\GalactoPINNS}{\package{GalactoPINNS}[https://github.com/charlottemyers/galactoPINNs]}

\newcommand{\PhiAB}{\phi_{\mrm{AB}}}
\newcommand{\PhiNN}{\tilde{\phi}_{\mrm{NN}}}

\newacronym{NODE}{NODE}{Neural Ordinary Differential Equation}
\newacronym{MW}{MW}{Milky Way}
\newacronym{LMC}{LMC}{Large Magellanic Cloud}
\newacronym[plural=BFEs, firstplural=basis function expansions (BFEs)]{BFE}{BFE}{basis function expansion}
\newacronym[plural=PINNs, firstplural=physics-informed neural networks (PINNs)]{PINN}{PINN}{physics-informed neural network}
\newacronym{BNN}{BNN}{Bayesian neural network}
\newacronym{MN}{MN}{Miyamoto--Nagai}
\newacronym{NFW}{NFW}{Navarro--Frenk--White}
\newacronym{AB}{AB}{analytic baseline}

\usetikzlibrary{calc,decorations.pathmorphing,decorations.pathreplacing,decorations.markings}
\tikzset{
  probwave/.style={
    draw,
    line width=0.4pt,
    decorate,
    decoration={coil, segment length=0.28cm, amplitude=0.5mm}
  }
}

\usetikzlibrary{calc,positioning}
\usetikzlibrary{calc,decorations.pathmorphing,decorations.pathreplacing,decorations.markings}
\tikzset{
  probwave/.style={
    draw,
    line width=0.4pt,
    decorate,
    decoration={coil, segment length=0.28cm, amplitude=0.5mm}
  }
}

\usetikzlibrary{arrows, decorations.markings,shapes,arrows,fit}
\tikzset{box/.style={draw, minimum size=2em, text width=8.9em, text centered},
bigbox/.style={
    draw,
    rectangle,
    minimum width=7.9cm,   
    minimum height=1.0cm,  
    inner sep=6pt
  },
lossbox/.style={
    draw,
    rectangle,
    minimum width=3.5cm,   
    minimum height=0.6cm,  
    inner sep=6pt
  }
}
\tikzset{
  boldarrow/.style={
    ->,
    line width=1.2pt,
    >={Latex[length=3.5mm, width=2.5mm]}
  }
}

\newcommand{\bigW}{8.4cm} 
\tikzset{
  bigboxfixed/.style={
    bigbox,
    text width=\bigW,
    align=left
  }
}

\begin{document}

\title{Reconstructing Galactic Gravitational Potentials from Stellar Kinematics with Physics-Informed Neural Networks}

\author[0009-0005-9329-8196]{Charlotte Myers}
\affiliation{Department of Physics and Kavli Institute for Astrophysics and Space Research, MIT, Cambridge, MA 02139, USA}
\email[show]{c\_myers@mit.edu}

\author[0000-0003-3954-3291]{Nathaniel Starkman}
\altaffiliation{Brinson Prize Fellow}
\affiliation{Department of Physics and Kavli Institute for Astrophysics and Space Research, MIT, Cambridge, MA 02139, USA}
\email{starkman@mit.edu}

\author[0000-0003-2806-1414]{Lina Necib}
\affiliation{Department of Physics and Kavli Institute for Astrophysics and Space Research, MIT, Cambridge, MA 02139, USA}
\affiliation{The NSF AI Institute for Artificial Intelligence and Fundamental Interactions, \\
Massachusetts Institute of Technology,\\
77 Massachusetts Avenue, Cambridge, MA 02139, USA}
\email{lnecib@mit.edu}

\begin{abstract} \label{sec:abstract}
    The gravitational potential of a galaxy encodes its mass distribution, formation history, and dark matter halo structure. %
    Accurate potential models are therefore critical for interpreting stellar kinematics, orbital dynamics, and the influence of satellite systems like the Large Magellanic Cloud. %
    Analytic potential models offer interpretability and efficiency but struggle to capture complex, non-axisymmetric structure and time-dependent 
    perturbations. %
    Neural network-based methods can capture this complexity but offer little interpretability.  %
    We introduce a physics-informed neural network (PINN) framework that combines data-driven learning with embedded physical constraints, available as the open-source package \GalactoPINNS. %
    Trained on acceleration measurements, the framework captures complex, small-scale features while preserving global physical consistency. %
    We test on systems of increasing complexity, from controlled analytic halos to cosmological simulations of Milky Way-like galaxies, achieving sub-percent acceleration errors with orbit reconstruction that consistently outperforms analytic baselines.  %
    Additionally, we implement a Bayesian neural network to provide spatially calibrated uncertainty estimates, and a time-dependent extension to capture smooth temporal evolution. %
    By treating an analytic model as a structured prior and learning corrections on top of it, the method retains physical interpretability while gaining the flexibility to represent realistic galactic potentials, making it well suited for Milky Way modeling and dynamical inference in the era of current and upcoming large-scale surveys.
\end{abstract}




\section{Introduction} \label{sec:intro}

    A galaxy's gravitational potential links the observed motions of tracers — stars, gas, and satellite galaxies — to the underlying distribution of mass \citep{Rubin:1980, BT08, BlandHawthorn:2016}. Accurately modeling this potential is therefore a central goal in galactic dynamics. %
    Because the majority of galactic mass resides in dark matter and interacts primarily gravitationally, potential modeling remains one of the primary probes of its spatial distribution and dynamical influence \citep[e.g.][]{Bertone:2005, McMillan:2017, STRIGARI20131}. %
    Gravitational potentials further underpin theoretical predictions for a wide range of dynamical observables, including stellar orbits, tidal streams, phase-space substructure, and secular evolution, making them indispensable for interpreting dynamics at the level of individual stellar data up to whole sky Galactic surveys \citep[e.g.][]{BT08, BlandHawthorn:2016, Watkins:2019, Bovy:2026}. %

    Beyond static field reconstruction, the long-term accuracy of a gravitational potential is essential for correctly modeling dynamical evolution. %
    Small, localized errors in the force field can lead to large deviations in reconstructed orbits, compromising our ability to reliably rewind stellar orbits to their origins and forward integrate streams and other dynamical tracers \citep{Merritt:1996, PriceWhelan:2016}. %
    This sensitivity places strong constraints on viable potential representations: they must be locally accurate, globally self-consistent, and stable under temporal integration across kpc scales and Gyr timescales. %
    These requirements are especially acute in the \gls{MW}, where long-standing work has demonstrated that the gravitational potential is neither static nor axisymmetric --- including but not limited to features like spiral arms, a rotating bar, and infalling and disrupting satellites \citep{Lin:1969:SpiralStructureDisk,Mathewson:1974:MagellanicStream,Ibata:1994:DwarfSatelliteGalaxy,Dwek:1995:MorphologyNearInfraredLuminosity,Drimmel:2001:ThreedimensionalStructureMilky,Belokurov:2006:FieldStreamsSagittarius,Wegg:2015:StructureMilkyWays,Helmi:2018:MergerThatLed,ou2025decodinggalactictwirldownfall}. %
    The infall of massive satellites such as the \gls{LMC} induces global, time-dependent perturbations in the \glsentrylong{MW} halo, producing non-inertial effects, large-scale wakes, and coherent motions in distant halo tracers \citep{Gomez2015, Cunningham_2020, Garavito-Camargo_2024, Vasiliev_2020, Erkal_2021, Erkal_2020, Shipp_2021, Pace_2022}. 
    More broadly, cosmological context and anisotropic mass assembly can imprint large-scale asymmetries on galactic potentials, further challenging idealized models \citep{Arora_2024}.

    Traditional approaches to gravitational potential modeling reflect a trade-off between interpretability, flexibility, and computational efficiency. %
    Analytic models employ physically motivated functional forms—such as NFW \citep{NFW:1997} or Einasto \citep{Einasto:1965} profiles—to describe dark matter halos and baryonic components, offering clear physical interpretation and efficient evaluation. %
    However, these models struggle to capture complex structure, including non-axisymmetry, localized perturbations, and departures from equilibrium \citep{ sands2024confrontingdiversityproblemlimits,ou2025decodinggalactictwirldownfall}. 
    More flexible representations, such as \glspl{BFE}, express the potential as a sum over orthogonal modes and have been widely applied to galactic systems \citep{Hernquist:1992:SelfconsistentFieldMethod,BT08, Arora_2024, Hunt_2025_SSA, Petersen_2025_EXP, panithanpaisal2025breakingtextsfcosmogemsmodelingunderstanding}. 
    \glspl{BFE} offer significant practical utility for N-body simulations, including $\mathcal{O}(N)$ computational scaling and natural parallelization \citep{Weinberg:1996:HighAccuracyMinimumRelaxation, Weinberg:1999:AdaptiveAlgorithmNBody}. %
    However, they often require a large number of modes to represent realistic potentials, can introduce unphysical features such as negative-density components, and become increasingly difficult to interpret as model complexity grows \citep[e.g.][]{Brown:1998:NbodySimulationsUsing, Wang:2020:BasisFunctionExpansions}. %
    Moreover, extending such approaches to time-dependent or strongly asymmetric systems remains challenging, though recent libraries have made progress in this direction more accessible \citep{Petersen:2022:EXPNbodyIntegration}. %

    Recent years have seen growing interest in machine-learning approaches to gravitational potential inference. %
    Several methods focus on recovering the potential statistically from fixed-time observations of phase-space data or kinematic tracers, emphasizing likelihood-based inference and uncertainty quantification \citep[e.g.][]{Green:2023:DeepPotential, Kalda2023, Buckley:2023:MeasuringGalacticDark, KaldaGreen2025, Lim:2025:MappingDarkMatter}. %
    More broadly, generative neural models offer flexible parameterizations of complex fields, though standard flow-based and optimal transport methods require careful treatment of conditional sampling to avoid training-testing misalignments \citep{Cheng:2025:CurseConditionsAnalyzing}. %
    While these techniques have achieved impressive results, many do not explicitly enforce the existence of a scalar potential or the exact force--potential relationship, and few are designed to guarantee consistency of long-term orbital dynamics under temporal integration. %

    At the same time, observational access to the Galactic acceleration field has expanded dramatically. %
    Stellar streams and other kinematic tracers have long provided localized constraints on the force field \citep[e.g.][and see  \citet{2023ARep...67..812B}]{Bovy+:2016, Monari+:2018, Malhan+Ibata:2019, Wegg+:2019, Posti+Helmi:2019, Ibata+:2024}. %
    More recently, direct acceleration measurements have become possible without relying on equilibrium assumptions or orbital models. %
    High-precision astrometry of quasars now yields the solar system's acceleration within the Milky Way 
    \citep{Bovy:2020:PurelyAccelerationbasedMeasurement}, while pulsar timing provides complementary constraints on vertical and in-plane force components across kiloparsec scales  \citep{Chakrabarti+:2021:MeasurementGalacticPlane, Moran+:2024:PulsarbasedMapGalactic, Donlon+:2025:WeighingMilkyWays}. %
    Stellar streams in particular encode three-dimensional acceleration along their tracks \citep{Nibauer2022}; recent analyses invert this signal into force-field maps and combine stream and pulsar constraints into joint inference \citep{Craig+:2023:BuildingAccelerationLadder}. %
    These datasets are heterogeneous in technique, spatial scale, and geometry. %
    Extracting coherent constraints on the gravitational potential from them requires a potential model that is flexible enough to fit diverse data, yet globally self-consistent. %
    
    Taken together, the limitations of other modeling approaches and the growing availability of acceleration data motivate a framework that embeds physical structure directly into a flexible, data-driven representation.
    \Glspl{PINN} provide such a framework: 
    rather than learning arbitrary mappings, \glspl{PINN} constrain the hypothesis space to functions that satisfy governing equations and structural relationships, such as the condition that forces derive from a scalar potential \citep{RAISSI2019686, cuomo2022scientificmachinelearningphysicsinformed, pinnsrecentadvances}. %
    In this sense, \glspl{PINN} can be viewed as a generalization of basis function expansions, learning adaptive, data-driven basis functions while retaining physical consistency. %
    By incorporating physical constraints into the loss function, \glspl{PINN} can reduce parameter counts, improve extrapolation, and mitigate degeneracies that arise in purely data-driven models \citep{Karniadakis2021, RAISSI2019686_pinns}. %
    Recent advances in physics-informed gravity modeling, notably the PINN-GM-III architecture \citep{pinngm}, have demonstrated state-of-the-art performance in reconstructing gravitational fields of terrestrial and small-body systems, achieving high accuracy and robust extrapolation beyond the training domain. %
    
    In this work, we extend the PINN-GM-III framework to the galactic regime, where larger spatial scales, multiple structural components, and pronounced asymmetries pose qualitatively new challenges. %
    We introduce a sequence of model variants that progressively embed physical structure, from basic Poisson consistency to multi-component decomposition and symmetry encoding, enabling the network to adapt to systems ranging from idealized triaxial halos to fully cosmological simulations. %
    Together, these developments yield a flexible yet physically grounded framework for modeling galactic gravitational potentials across a wide range of complexity and dynamical states. %
    This paper is organized as follows. 
    In \autoref{sec:methods}, we describe the framework and the physical reasoning behind 
    each of its components, introducing a sequence of model variants that build on one 
    another by progressively adding more physical structure. %
    In \autoref{sec:results}, we evaluate these models on a sequence of increasingly realistic test systems, from a controlled triaxial halo to a realistic cosmological simulation of the Milky Way.
    In \autoref{sec:discussion}, we examine the broader implications of these results, including limitations of the current formulation and a discussion of extensions toward observational data.


\section{Methods} \label{sec:methods}

    \input{figure_schematic}

    In this work, we aim to reconstruct galactic gravitational potentials by combining the interpretability of analytic models with the flexibility of neural networks. %
    Concretely, we start from a physically-motivated analytic model that captures the dominant large-scale structure of the potential, and train a neural network to learn the residual -- whatever structure the analytic model does not capture. %

    The framework operates in two phases. %
    During training, the model learns to reproduce the gravitational field from a set of paired positions $\mathbf{x}$ and gravitational accelerations $\mathbf{a}$. %
    The true potential itself is never required as an input; it is an output of the model, inferred from the accelerations through the physical relation $\mathbf{a} = -\nabla\phi$. %
    In principle, the training data can come from analytic models, $N$-body simulations, or observational kinematic measurements, making the framework applicable in a wide range of settings. %
    During evaluation, the trained model can predict the potential and acceleration anywhere in space, not just where training data were available. %
  
    \autoref{fig:schematic} illustrates the training phase, which we briefly walk through here. %
    Starting from the left, a 3D snapshot of the target potential is used to extract particle positions $\mathbf{x}(t)$ and their corresponding gravitational accelerations $\mathbf{a}(t)$. %
    In the figure, these are drawn from an analytic Milky Way model introduced in \autoref{sec:results:static_mw_lmc}. %
    The target potential itself is only needed during training to generate these labeled $(\mbf{x}, \mbf{a})$ pairs; at inference time, the model requires only $(\mbf{x}, \mbf{a})$, not the potential. %
    The positions are passed through a coordinate transformation (top center) to align more naturally with the geometry of galactic potentials. %
    This transformation is discussed in detail in \autoref{sec:methods:input_representation}. %
    Two physics-informed priors -- a scaling function $n(\mathbf{x})$ (see \autoref{sec:methods:radial_scaling}) and an \acrfull{AB} $\phi_\mathrm{AB}$ (see \autoref{sec:methods:analytic_fusing}) -- are then combined with the output of a neural network to rebuild the full potential $\phi$ (center right). %
    Together, these choices free the network from learning well-understood large-scale structure, directing its capacity towards features that are genuinely hard to model analytically. %

    Finally, we differentiate the reconstructed potential and compare it to the true accelerations through the loss function, as shown in the bottom right of \autoref{fig:schematic}. %
    We drive the network towards physically consistent solutions through two components: a primary term that enforces the conservative force law between the learned potential and the known accelerations (see \autoref{sec:methods:loss}), and an optional term that encourages long-term orbit energy conservation (see \autoref{sec:methods:orbit_energy}). %

    When multiple temporal snapshots are available, we extend the framework to time-evolving potentials: rather than treating each snapshot independently, we train the network to learn how the potential changes over time and integrate these changes continuously between snapshots, as described in \autoref{sec:methods:modeling_time_dependence}. %
    As shown in \autoref{fig:schematic}, we pass time $t$ alongside the positions into the coordinate transformation, and reconstruct the potential as an integral over the learned time derivatives (center right). %

    In practice, we present our framework through a set of modular design choices that can be enabled or disabled depending on the application, and which can be easily applied using our open source code \GalactoPINNS. %
    In \autoref{sec:methods:variants}, we therefore introduce several model variants that sequentially incorporate these features. %
    In the remainder of \autoref{sec:methods}, we discuss the motivation and implementation of each element in detail. %

    \subsection{Model variants (PINN I--VI)}\label{sec:methods:variants}

        To assess the individual contribution of each model component, we build a sequence of progressively more expressive variants, labeled \textbf{PINN~I} through \textbf{PINN~VI}.  %
        All variants share the same core architecture, but differ in which additional physical structure is incorporated into the model. %
        Concretely, each variant builds on a previous one by activating one additional feature (i.e. a modified coordinate system) that is designed to improve either accuracy, physical consistency, or interpretability. %
        This cumulative design allows the contribution of each individual feature to be isolated and evaluated independently. %

        \vspace{10pt}
        
        \noindent\textbf{PINN Variants}\vspace{-4pt}
        \begin{enumerate}[label=\textbf{\Roman*.}, ref=\Roman*, leftmargin=1.5em]
            \item \label{model:pinnI} %
                \textbf{Baseline:} A neural network predicts the gravitational potential and is trained by enforcing the known relation between the potential and  acceleration ($\mathbf{a} = -\nabla\phi$) through the loss function (see \autoref{sec:methods:loss}). %
                An optional energy-conservation term can be included to further improve orbit accuracy (see \autoref{sec:methods:orbit_energy}). %
            \item \label{model:pinnII} %
                \textbf{(I) + spherical input representation:}  Rather than feeding the network raw Cartesian positions $(x, y, z)$, we instead use a spherical coordinate system that is better suited to the geometry of galactic potentials and remains well-behaved at both small and large radii (see \autoref{sec:methods:input_representation}). %
            \item \label{model:pinnIII} %
                \textbf{(II) + spatial scaling:} 
                 The potential is divided by a physically motivated scaling function $n(\mathbf{x})$ before being passed to the network. %
                 $n(\mathbf{x})$ removes the dominant large-scale structure which is already well understood analytically so the network can focus on learning the smaller, more complex deviations from it (see \autoref{sec:methods:radial_scaling}). %
            \item \label{model:pinnIV} %
                \textbf{(III) + \acrlong{AB}:} A physically motivated mass model (e.g., an NFW halo) is incorporated directly into the reconstruction, so the network only needs to learn departures from this known large-scale structure rather than the full potential from scratch (see \autoref{sec:methods:analytic_fusing}). %
            \item \label{model:pinnV} %
                \textbf{(IV) + trainable analytic parameters:} Extends PINN~\ref{model:pinnIV} by not fixing the parameters of the analytic mass model (e.g., halo mass, scale radius) in advance. %
                Instead, the parameters are fitted jointly with the neural network during training, allowing the network to correct for uncertainty in the assumed mass model (see \autoref{sec:methods:trainable_analytic_parameters}).
            \item \label{model:pinnVI} %
                \textbf{(IV/V) + time dependence:} Extends the framework to time-evolving potentials by learning how the potential changes with time, rather than treating each snapshot independently. %
                This enforces a smooth, physically consistent temporal evolution across snapshots (see \autoref{sec:methods:modeling_time_dependence}). %
        \end{enumerate}
        While we evaluate PINN~\ref{model:pinnIII} on a simple triaxial test system in \autoref{sec:results:triaxial_nfw}, PINN~\ref{model:pinnIV} represents the minimum configuration suitable for most realistic systems. %
        Additionally, all variants can be paired with a Bayesian inference scheme, as described in \autoref{sec:methods:bayesian}. %
        Rather than producing fixed network weights, the Bayesian treatment represents weights and analytic parameters as probability distributions, yielding spatially varying, calibrated uncertainty estimates on the reconstructed potential and forces. %
        We denote Bayesian variants with a \textbf{B} suffix (e.g., PINN~\ref{model:pinnV}-B denotes a PINN~\ref{model:pinnV} model under Bayesian inference). %

        We provide detailed descriptions of each model variant in the remainder of \autoref{sec:methods}; readers primarily interested in the results may skip ahead to \autoref{sec:results}. %


    \subsection{PINN I: Physics-Informed Loss Function}\label{sec:methods:loss}
        Our goal is to reconstruct the gravitational potential $\phi(\mbf{x})$ of a galactic system from a set of acceleration measurements $\mbf{a}(\mbf{x})$ -- the gravitational force per unit mass at positions $\mbf{x}$ sampled from the target field. %
        Rather than learning the accelerations directly, we instead train a neural network to predict a scalar potential $\phi$, and define a loss on its gradient $\nabla\phi$ using the fundamental relation $\mbf{a}=-\nabla\phi$. %

        The baseline model accepts \textit{N} paired vectors of positions and accelerations, $(\mbf{x}_i, \mbf{a}_i)$, sampled from the target gravitational field. %
        In this work, these training pairs are obtained either by evaluating analytic potential models on a grid, or by sampling accelerations from $N$-body simulations; in both cases the accelerations are treated as noiseless. %
        
        The network, parameterized by $\mbs{\theta}$, outputs the predicted potential $\phi(\mbf{x}_i |\mbs{\theta})$, which is differentiated with respect to position $\mbf{x}$ to produce an acceleration vector $\mbf{a}$. %
        The physics-informed loss $\mcal{L}_{\text{acc}}$ enforces the constraint $\mbf{a} = -\nabla\phi$ by combining absolute and relative acceleration errors: %
        \begin{align}\label{eq:model_design:architecture:loss}
            \mathcal{L}_{\text{acc}}(\boldsymbol{\theta})
            &= \frac{1}{N} \sum_{i=1}^{N} \Bigg(
                \left\| -\nabla \phi(\mathbf{x}_i \mid \boldsymbol{\theta}) - \mathbf{a}_i \right\| \\
            &\qquad
                + \lambda_{\mathrm{r}}
                \frac{
                    \left\| -\nabla \phi(\mathbf{x}_i \mid \boldsymbol{\theta}) - \mathbf{a}_i \right\|
                }{
                    \left\| \mathbf{a}_i \right\| + \epsilon_{\mathrm{mach}}
                }
            \Bigg) \; .
        \end{align}
        The relative term, weighted by $\lambda_\text{r}$, prevents loss of accuracy at large radii where accelerations decay toward zero. %
        $\epsilon_{\mathrm{mach}}$ is the machine epsilon, to avoid division by zero at locations where the acceleration vanishes, such as the galactic center or in the far field. %

        Because the loss acts on the gradient of the predicted potential rather than on accelerations directly, every optimization step enforces the relation  $\mbf{a} = -\nabla\phi$, tying the learned field to physically consistent solutions. %
        This corresponds to the ``Autodiff.\ + Loss'' block in \autoref{fig:schematic} (bottom right), where the reconstructed potential is differentiated and compared to the true accelerations $\mbf{a}(t)$ from the snapshot. %


    \subsection{PINN II: (I) + Spherical input representation}\label{sec:methods:input_representation}

        Rather than using Cartesian coordinates directly, we pass positions through the five-dimensional spherical coordinate transformation of \citet{pinngm} before they enter the network, as shown as the ``5D spherical coords.'' block in \autoref{fig:schematic}. %
         
        In this parameterization, two compactified radial coordinates $(r_i,r_e)$ map the interior and exterior regions to finite intervals. %
        This transformation eliminates the large dynamic range that would arise from passing raw distances to the network. %
        Additionally, this representation is natural to the geometry of galactic potentials, which are broadly organized around a central mass concentration with approximate radial symmetry. %

        We define three angular coordinates $(s,t,u)$ to describe direction without the standard singularities of spherical coordinates, as shown in \autoref{sec:appendix:5d_coords}. %
        This keeps both the potential $\phi$ and its spatial derivatives well-conditioned across all spatial scales. %
        As a result, the network is encouraged to reconstruct the potential accurately at all radii, not just in regions that dominate the raw numerical scale of the inputs. %

        In addition to the change in coordinate system, all input variables are non-dimensionalized using physically consistent scaling factors to ensure that positions, accelerations, and potentials remain numerically comparable; details of the non-dimensionalization procedure are provided in \autoref{sec:appendix:scaling}. %


    \subsection{PINN III: (II) + Spatial scaling}\label{sec:methods:radial_scaling}

        Gravitational potentials decay toward zero at large radii, producing near-zero target values that can lead to numerical underflow and degrade extrapolation. %
        Since this radial trend is well described by analytic forms, we factor it out explicitly using a physically-motivated scaling function $n(\mbf{x})$. %
        The choice of $n(\mbf{x})$ encodes our prior expectation of the dominant spatial structure of the potential -- it is one of the two ``physics-informed priors'' shown in \autoref{fig:schematic} that reduce the complexity of what the network must learn. %
        
        Expanding on the design of \citet{pinngm}, we factor out $n(\mbf{x})$ from the full predicted potential $\phi(\mbf{x}|\mbs{\theta})$, and train the network to predict a scaled potential $\PhiNN(\mbf{x}|\mbs{\theta}) = \phi(\mbf{x}|\mbs{\theta}) \cdot n(\mbf{x})$, where the full potential is reconstructed as:
        \begin{equation}\label{eq:phi_scaled}
            \phi(\mbf{x}|\mbs{\theta}) = \frac{1}{n(\mbf{x})} \PhiNN(\mbf{x}|\mbs{\theta})  \, .
        \end{equation}
        The scaling function \(n(\mathbf{x})\) is chosen to approximate the leading-order shape of the potential while remaining smooth and well behaved across the domain. %
        This compresses the dynamic range of the output, improving numerical conditioning of both $\phi$ and $\nabla\phi$ and directing the loss in \autoref{eq:model_design:architecture:loss} toward structure that deviates from the dominant radial fall-off. %
        
        In contrast to \citet{pinngm}, who use a purely radial scaling \(n(r)\), we adopt a more general spatially dependent function \(n(\mathbf{x})\). %
        This extension is motivated by the structure of galactic potentials, which often include strongly non-spherical components such as bars, disks, and satellite perturbations. %
        While radial scaling captures the overall halo trend, it cannot account for directional variations introduced by these components. %
        Allowing the scaling function to depend on the full spatial coordinate therefore provides additional flexibility while preserving the primary goal of the transformation: removing the dominant large-scale variation so that the neural network focuses on higher-order structure. %

        As a toy example: suppose the true potential is a triaxial NFW halo with scale radius $r_s$ and mass scale $M$. %
        In this case, the dominant spatial dependence can be written in terms of the ellipsoidal radius
        \begin{equation} \label{eq:ellipsoid-r}
            R_{\mathrm{ell}}(\mathbf{x})
                =
                \sqrt{
                x^2
                +
                \frac{y^2}{q_1^2}
                +
                \frac{z^2}{q_2^2}
                },
        \end{equation}
        where $q_1$ and $q_2$ denote axis ratios relative to the $x$-axis.
        The true potential then takes the form

        \begin{equation}
        \label{eq:trixial_nfw}
            \phi(\mathbf{x})
            =
            -\frac{GM}{R_{\mathrm{ell}}(\mathbf{x})}
            \ln\!\left(1+\frac{R_{\mathrm{ell}}(\mathbf{x})}{r_s}\right).
        \end{equation}
        
        
        Choosing the scaling function
        \begin{equation} \label{eq:radial_scaling_trixianl_nfw}
            n(\mathbf{x})
            =
            \frac{R_{\mathrm{ell}}(\mathbf{x})}
            {r_s
            \ln\!\left(1+\dfrac{R_{\mathrm{ell}}(\mathbf{x})}{r_s}\right)}
        \end{equation}
        factors out the full spatial dependence of the potential, such that the optimal network prediction is simply
        \begin{equation}
            \PhiNN(\mathbf{x}|\mbs{\theta}) = -\,\frac{GM}{r_s}.
        \end{equation}

        In this idealized case, the scaling function absorbs the entire structure of the potential, leaving the network to learn only the overall normalization. %
        In more realistic settings, the true potential is not known and the scaling function only approximates the dominant geometry. %
        Yet even an imperfect choice of $n(\mbf{x})$ can be beneficial, as any structure it absorbs is structure the network no longer needs to learn. %
        The scaling function acts as a regularizer in function space, compressing the dynamic range of $\PhiNN$, concentrating the network toward solutions consistent with the physical prior, and preserving its expressive capacity for higher-order features. %

        
    \subsection{PINN IV: (III) + Analytic baseline}\label{sec:methods:analytic_fusing}

        In many astrophysical settings, the large-scale structure of the gravitational potential is already well-constrained by analytic models, which capture the dominant mass distribution of a galaxy to reasonable accuracy. %
        Rather than asking the network to re-learn these well-understood features from scratch, we incorporate an \acrlong{AB} potential $\PhiAB(\mbf{x})$ directly into the model architecture. %
        The network then predicts only the scaled residual field $\PhiNN(\mbf{x}|\mbs{\theta})$, which is typically smaller in amplitude and more localized in structure, and therefore easier to learn accurately. %
        Modifying \autoref{eq:phi_scaled}, the full potential is then reconstructed as:  %
        \begin{equation}\label{eq:phi_fused}
            \phi(\mbf{x}|\mbs{\theta}) = \PhiAB(\mbf{x}) + \frac{1}{n(\mbf{x})} \PhiNN(\mbf{x}|\mbs{\theta})  \, .
        \end{equation}
        This fusion leverages the interpretability and reliability of analytic modeling while reserving the network's flexibility for capturing higher-order structure such as non-axisymmetric perturbations. %

        Like the scaling function $n(\mbf{x})$, the \acrshort{AB} is chosen based on prior knowledge of the target potential, so together they form the ``physics-informed priors" shown in the center block of \autoref{fig:schematic}. Setting \(\PhiAB\!\equiv\!0\) recovers the scaling-only variant from \autoref{sec:methods:radial_scaling}. %

        Again, consider a toy example in which the true potential is a triaxial NFW halo (e.g. \autoref{eq:trixial_nfw}). %
        If the baseline is chosen to be a spherical NFW halo with the same mass and scale radius,
        \begin{equation} \nonumber
            \PhiAB = \phi_{\rm{NFW}}(r) =
            -\,\frac{GM}{r}
            \ln\!\left(1 + \dfrac{r}{r_s}\right),
        \end{equation}
        then the fused model in Eq.~\eqref{eq:phi_fused} becomes
        \begin{equation} \nonumber
            \phi(\mathbf{x}|\mbs{\theta}) =
            \phi_{\rm{NFW}}(r) + \frac{1}{n(\mathbf{x})}\PhiNN(\mbf{x}|\mbs{\theta}).
        \end{equation}
        In this case, the network learns the deviation between the spherical and triaxial profiles, focusing only on the correction associated with the triaxiality. 

        In this example we have not explicitly chosen $n(\mbf{x})$, though \autoref{eq:radial_scaling_trixianl_nfw} would be appropriate. %
        A natural choice for the scaling function is $n(\mathbf{x}) = 1/\PhiAB(\mathbf{x})$, since the analytic baseline already captures the dominant spatial scaling of the potential. %
        While not required, this choice allows \autoref{eq:phi_fused} to be written in a more compact form,
        \begin{equation}\label{eq:phi_fused_condensed}
            \phi(\mathbf{x}|\mbs{\theta}) = \PhiAB(\mathbf{x}) \left(1 + \PhiNN(\mathbf{x}\,|\,\boldsymbol{\theta})\right).
        \end{equation}
        
        Under this parameterization, the neural network predicts the fractional deviation from the analytic baseline rather than the absolute residual potential. %
        This interpretation can be particularly convenient when the analytic model provides a good first-order description of the system. %
        In \autoref{sec:results:triaxial_nfw}, we adopt this formulation to model a triaxial halo, using a spherical NFW as both the analytic baseline and the scaling function $n(\mathbf{x})$. %

     \subsection{PINN V: (IV) + Trainable analytic parameters} \label{sec:methods:trainable_analytic_parameters}
    
        In realistic applications, the parameters of the analytic baseline are rarely known with high precision; for instance, the mass and scale radius of a dark matter halo inferred 
        from stellar kinematics carry significant uncertainty. %
        Rather than fixing these parameters before training, we allow them to be optimized jointly with the neural-network weights within a single model. %
        This is reflected in \autoref{fig:schematic}, where $\PhiAB$ (center right) is itself a learned component (parameterized by $\mbs{\theta}_3$) which is combined with the learned residual to reconstruct the full potential. %
        Within this framework, the large-scale structure and higher-order deviations can be adjusted together in a self-consistent manner. %
        
        When combined with spatial scaling (as discussed in \autoref{sec:methods:radial_scaling}), we also enforce consistency between the baseline model and the $n(\mbf{x})$ function. %
        In particular, when $n(\mbf{x})$ depends on the halo scale radius $r_s$, treating $r_s$ as a trainable parameter allows the scaling to remain aligned with the evolving baseline 
        throughout training, rather than being fixed to a potentially misspecified prior value. %

        The training strategy can be adapted depending on whether the primary goal is accurate field reconstruction or interpretable parameter inference. %
        In a fully joint training scheme, the analytic parameters and neural-network weights are optimized simultaneously. %
        This approach captures correlations between the analytic and residual components but can introduce additional optimization complexity. %
        Alternatively, a sequential strategy can first optimize the analytic baseline to capture the dominant large-scale structure and then train the residual to model higher-order corrections. %
        Such staged training can improve stability and yield analytic parameters that more closely reflect the underlying physical system. %
        In practice, the framework introduced here supports either strategy, as well as intermediate approaches that gradually transition from analytic-dominated to residual-dominated optimization. %
        
    

    \subsection{PINN~VI: (IV/V) + Modeling time dependence}\label{sec:methods:modeling_time_dependence}

       The design features introduced thus far were presented for the static potential case, but remain applicable when the field evolves in time. %
       For time-dependent applications, we assume that multiple snapshots are available, each providing positions and accelerations at discrete times $t_1, \dots, t_n$. %
       The goal is then to reconstruct a continuous, smoothly evolving potential between these snapshots, rather than treating each one independently. %
       
       To achieve this, we introduce a neural integral formulation, illustrated in the center right of \autoref{fig:schematic}. %
       The full time-dependent potential is built from three learned components -- an initial spatial correction $\tilde{\phi}_{\mathrm{NN}_0}$ (parameterized by $\mbs{\theta}_1$), a learned time derivative $\dot{\tilde{\phi}}_\mathrm{NN}$ (parameterized by  $\mbs{\theta}_2$), and the \acrshort{AB} $\phi_\mathrm{AB}$ (parameterized by $\mbs{\theta}_3$ when trainable, as discussed in \autoref{sec:methods:trainable_analytic_parameters}). %
        The full potential is reconstructed as:
        \begin{align}\label{eq:node}
            \phi(\mathbf{x}, t \!\mid\! \boldsymbol{\theta})
            &= \phi_{\mathrm{AB}}(\mathbf{x}, t \!\mid\! \mbs{\theta}_3)
            + \frac{\tilde{\phi}_\mathrm{NN}(\mathbf{x}, t \!\mid\! \boldsymbol{\theta})}{n(\mathbf{x}, t)}
            \\ \intertext{where} 
            \tilde{\phi}_\mathrm{NN}(\mathbf{x}, t \!\mid\! \boldsymbol{\theta})
            &= \tilde{\phi}_{\mathrm{NN}_0}(\mathbf{x} \!\mid\! \boldsymbol{\theta}_1)
            + \int_{0}^{t}\! \dot{\tilde{\phi}}_\mathrm{NN}(\mathbf{x}, t' \!\mid\! \boldsymbol{\theta}_2)\,
                \mathrm{d}t'
        \end{align}

        This formula extends naturally from \autoref{eq:phi_fused}, with the neural residual now comprising an initial correction $\tilde{\phi}_{\mathrm{NN}_0}$ at $t = 0$ plus an integrated correction that accumulates the learned time derivative $\dot{\tilde{\phi}}_\mathrm{NN}$ up to the evaluation time $t$. The integral is evaluated numerically via 3-point Gauss--Legendre quadrature \citep{Gauss:1814:MethodusNovaIntegralium}. %

        A simple approach to modeling temporal variation is to append time as an additional input coordinate to the network. %
        However, this treatment regards time as an unordered feature and imposes no causal or structural constraint on the learned evolution. %
        As a result, the model must independently learn the field at each time, with no guarantee of temporal consistency. %
        By contrast, the neural-integration formulation constrains the evolution to follow a continuous trajectory by learning the time derivative of the residual field and integrating it. %
        This construction enforces smooth evolution by design and couples predictions across time through the integration. %
        
        Although learning a derivative can, in principle, be more challenging than learning the potential directly, since derivatives may exhibit sharper local structure, in this setting the derivative is supervised only through its integrated effect. %
        The integration step effectively smooths high-frequency components, making the optimization more stable. %
        In addition, the constrained temporal structure reduces the representational burden on the network, allowing accurate time-dependent modeling with a smaller architecture than would be required by treating time as an unconstrained input coordinate. %


    \subsection{PINN-B Variants: Bayesian networks }\label{sec:methods:bayesian}
        
        Accurate reconstruction of the potential is necessary but not sufficient for reliable inference --- it is equally important to know where the model's predictions can be trusted and where they are poorly constrained by the available data. %
        To quantify predictive uncertainty, we adopt a \gls{BNN} framework in which network weights and analytic parameters are treated as random variables rather than fixed point estimates \citep{bnn2}. %
        This is reflected in the top panel of \autoref{fig:schematic}, where each set of parameters $\mbs{\theta}$ is sampled from a learned probability distribution rather than taking a single fixed value. %
        While all distributions used in this work are Gaussian, in principle one could sample parameters from any distribution that suits their application. %
        By drawing multiple samples from these distributions, we obtain an ensemble of potential reconstructions whose spread provides a spatially varying uncertainty estimate --- larger where the model is weakly constrained by the training data, and smaller where it is well-determined. %

        This uncertainty can be propagated directly into physical quantities such as reconstructed forces and integrated orbits, providing a principled indicator of where predictions should and should not be trusted. %
        The framework is flexible in terms of which parameters are treated probabilistically: in the full time-dependent model all three parameter sets ($\mbs{\theta}_1$, $\mbs{\theta}_2$, $\mbs{\theta}_3$) are inferred jointly, while in the static case without a trainable baseline only $\mbs{\theta}_1$ is needed. %
        In all cases, the same inference machinery applies. %

        We implement this framework in \GalactoPINNS{} using \texttt{NumPyro} \citep{numpyro2019} and approximate the posterior distributions via stochastic variational inference (SVI) \citep{SVIhoffman2013stochasticvariationalinference, SVIwingate2013automatedvariationalinferenceprobabilistic}. %
        We place truncated-normal priors (with variance $\sigma_\alpha^2$) on the analytic parameters $\mbs{\alpha}$ within physically plausible ranges, and zero-mean Gaussian priors with variance $\sigma_\theta^2$ on the neural network weights $\mbs{\theta}$. %
        Accelerations are modeled with a Gaussian likelihood
        \begin{equation} \label{eq:posterior}
            p(\mbf{a}_{\mathrm{obs}} \mid \mbf{x}, \mbs{\alpha}, \mbs{\theta}) = 
            \mathcal{N}\!\big(\mbf{a}_{\mathrm{pred}}(\mbf{x}; \mbs{\alpha}, \mbs{\theta}),\, 
            \sigma_a^2\big),
        \end{equation}
        with fixed noise $\sigma_a = 2\times10^{-4}$, chosen to be small relative to the typical acceleration magnitudes in our training data; in principle, this can be adapted to match the noise level of any target dataset. %
        The variational posterior $q(\mbs{\alpha},\mbs{\theta})$ is a fully factorized Gaussian (\texttt{AutoNormal}) optimized by maximizing the ELBO \citep{kingma2022autoencodingvariationalbayes, Jordan1998}. %
        The physics-informed loss (\autoref{eq:model_design:architecture:loss}) 
        enters naturally through the likelihood term, so the variational objective jointly enforces physical consistency and posterior calibration. %
            
        To stabilize joint inference of the analytic and residual components, we generally adopt the two-stage training strategy described in \autoref{sec:methods:trainable_analytic_parameters}: an initial phase with a broad weight prior fits the analytic baseline to the dominant large-scale structure, followed by a phase with tighter analytic priors and relaxed neural priors that allows the residual network to capture higher-order corrections. %
        Each stage is initialized from the previous variational solution to ensure continuity in the posterior evolution. %
        Specific prior values and training details are provided in \autoref{sec:appendix:training_details}. %

    \subsection{Enforcing energy conservation}\label{sec:methods:orbit_energy}

        Pointwise agreement between predicted and true accelerations is necessary but not sufficient to guarantee physically consistent dynamics \citep{Dehnen+Read:2011, Wang:2020:BasisFunctionExpansions}. %
        Small, spatially correlated force errors can accumulate over time and lead to secular energy drift and incorrect orbital dynamics \citep{Wisdom+Holman:1991:Symplectic, Hut+Makino+McMillan:1995:Leapfrog}. %
        
        To promote dynamical consistency, we introduce an optional term to the loss function that penalizes violations of energy conservation along test-particle orbits. %
        We emphasize that this formulation assumes a time-independent potential, since energy conservation in the form $\dot{E} = 0$ holds only when $\partial_t \phi = 0$. This term is therefore applied only to static reconstructions, or on a fixed-time-per-snapshot basis in the time-dependent settings of \autoref{sec:methods:modeling_time_dependence}. %

        We consider a set of pre-computed orbit trajectories $\{\mathbf{x}_k(t), \mathbf{p}_k(t)\}$ generated in the true gravitational potential. %
        For computational efficiency, the orbits are pre-computed prior to training. %
        For a reconstructed potential $\phi(\mathbf{x}\mid \boldsymbol{\theta})$, the implied specific energy along a trajectory is %
        \begin{equation}\label{eq:energy_pred}
            E_k(t\mid \boldsymbol{\theta})
            \;=\;
            \frac{1}{2}\,\|\mathbf{p}_k(t)\|^2
            \;+\;
            \phi\!\left(\mathbf{x}_k(t)\mid \boldsymbol{\theta}\right),
        \end{equation}
        where $\mathbf{p}_k(t)$ denotes the particle momentum (in units with $m=1$). %
        In a time-independent potential, the true specific energy $E_k$ is constant along each orbit; deviations of ${E}_k(t)$ from a constant therefore provide a probe of global inconsistencies in the reconstructed potential. %

        Rather than comparing the predicted energy  \({E}_k(t\mid\boldsymbol{\theta})\) to an external reference energy, we measure self-consistency of the energy along each trajectory. %
        Specifically, we penalize the fractional deviation of ${E}_k(t\mid \boldsymbol{\theta})$ from its value at the initial time $t_0$: %
        \begin{equation} \label{eq:energy_loss}
            \mathcal{L}_{\mathrm{E}}(\boldsymbol{\theta})
            \;=\;
            \frac{1}{N_{\mathrm{orb}}}
            \sum_{k=1}^{N_{\mathrm{orb}}}
            \frac{1}{T}
            \sum_{t}
            \left(
                \frac{{E}_k(t\mid \boldsymbol{\theta}) - {E}_k(t_0\mid \boldsymbol{\theta})}
                    {\left|E_k(t_0\mid \boldsymbol{\theta})\right| + \epsilon}
            \right)^{\!2},
        \end{equation}
        where $N_{\mathrm{orb}}$ is the number of orbit trajectories, $T$ is the number of time steps, and $\epsilon$ is a small constant for numerical stability. %

        Although the loss is expressed as a deviation from $t_0$, this choice does not privilege the initial point in any meaningful sense. %
        The loss is zero if and only if ${E}_k(t\mid \boldsymbol{\theta})$ is constant along the entire trajectory, which is the correct physical condition regardless of which reference time is chosen. %
        Because ${E}_k(t_0\mid \boldsymbol{\theta})$ appears only in the denominator as a fixed normalization scale (via $\big|{E}_k(t_0\mid \boldsymbol{\theta})\big|$), the gradient signal is dimensionless and comparable across orbits with very different energy scales. %

        The acceleration-based loss in \autoref{eq:model_design:architecture:loss} is augmented with this energy-conservation term as %
        \begin{equation}
         \label{eq:combined_loss}   \mathcal{L}_{\mathrm{total}}(\boldsymbol{\theta})
            \;=\;
            \mathcal{L}_{\mathrm{acc}}(\boldsymbol{\theta})
            \;+\;
            \lambda_{\mathrm{E}}(s)\,\mathcal{L}_{\mathrm{E}}(\boldsymbol{\theta}),
        \end{equation}
        where the gradient contributions of the two terms are rescaled to have equal $L^2$ norm before being combined, preventing either term from dominating purely due to differences in loss magnitude. %

        To stabilize early training, $\lambda_{\mathrm{E}}$ is not applied at its full value from the start. %
        Instead, it is ramped from zero to its target value over the course of training according to a cosine schedule, %
        \begin{equation}
            \lambda_{\mathrm{E}}(s)
            \;=\;
            \lambda_{\mathrm{E}}^{\max}
            \cdot
            \frac{1 - \cos\!\left(\pi\, s / S\right)}{2},
        \end{equation}
        where $s$ is the current training step and $S$ is the total number of steps. %
        This allows the acceleration loss to establish a reasonable initial potential before the energy-conservation term begins to exert strong influence. %
        
        In practice, the energy-conservation term is most effective when used as a weak regularizer, complementing the local acceleration loss by suppressing non-physical distortions of the potential. %
        Although both loss terms constrain the same underlying field (since energy conservation follows directly from the force law), they are not redundant in finite-sample settings: the acceleration loss enforces local consistency, while the orbit-based energy term aggregates these constraints along extended trajectories, amplifying small, spatially correlated errors that may be difficult to detect pointwise. %

        This loss does not require differentiating through orbit integration and therefore adds negligible computational cost. Instead, it uses pre-computed trajectories as global probes of the reconstructed potential. %
        Hence, this approach is well suited to simulation-based applications, where particle tracks or high-fidelity \textit{N}-body trajectories are readily available from the simulation. %
        
        Alternatively, instead of pre-computing orbits, one can integrate short test orbits directly within the learned potential at each optimization step, and check that energy is conserved. %
        This adds some computational cost during training but remains manageable and provides a direct test of whether the learned potential is energy-conserving. %


    
      \subsection{Gauge Freedom}\label{sec:methods:boundary_enforcement}

        Because the training objective is formulated in terms of accelerations, the learned potential is only identifiable up to an additive constant -- a gauge freedom that is inherent to any acceleration-based inference. In the time-dependent setting, this freedom is further enlarged: a constant offset in the learned time derivative $\dot{\tilde{\phi}}_\mathrm{NN}$ integrates to an additional time-dependent offset in $\phi(\mathbf{x}, t)$. While such offsets have no effect on the forces or orbits, they complicate direct comparison of the learned potential against a ground truth.
        
        One way to fix this gauge is to enforce a transition to the analytic baseline at large radii, anchoring the learned potential to a known reference in the outer regions where the analytic model is most reliable. This strategy was also adopted by \citet{pinngm} for terrestrial systems, and can in principle improve far-field reconstruction. However, as noted there, enforcing a boundary transition is not always appropriate -- particularly for time-dependent potentials where the outer field may evolve due to infalling satellites or other large-scale perturbations. We therefore do not apply a boundary transition in any of the experiments presented in this paper.
        
        Instead, we evaluate potential-level errors using gauge-fixed quantities. Specifically, we compare potential differences relative to a reference point,
            \begin{equation}
                \Delta\phi(\mathbf{x}, t)
                \;=\;
                \phi(\mathbf{x}, t)
                -
                \phi(\mathbf{x}_{\mathrm{ref}}, t),
            \end{equation}
        which are invariant under additive gauge transformations. Unless otherwise stated, the reference point $\mathbf{x}_{\mathrm{ref}}$ is chosen in the far field ($r > 200 \ \mathrm{kpc}$), where 
        the potential is well described by smooth analytic models and the gauge offset is effectively absorbed. All reported potential residuals are computed using $\Delta\phi$; force- and orbit-level diagnostics are fully gauge invariant by construction and require no such correction.


\section{Results} \label{sec:results}

    To evaluate the framework introduced in \autoref{sec:methods}, we apply it to a sequence of test systems of increasing physical complexity. %
    We build up the assessment gradually, starting from simple, well-understood systems where we can isolate the effect of individual design choices, and progressing toward a full cosmological simulation where the gravitational field reflects the complexity of a realistic galaxy. %

    In \autoref{sec:results:triaxial_nfw} and \autoref{sec:results:static_mw_lmc}, we consider static systems built entirely from analytic components, where the ground-truth potential is known exactly. %
    We use analytic baselines that are intentionally misaligned with the true potentials (missing components, using incorrect parameters, or both) to test whether the \gls{PINN} framework can absorb model misspecification through the learned residual. %
    In \autoref{sec:results:triaxial_nfw}, we begin with a simple triaxial NFW halo to isolate the effects of radial scaling and analytic fusion (PINN~\ref{model:pinnIII} and PINN~\ref{model:pinnIV}) in a controlled setting. %
    This minimal example allows us to assess the residual learning without the added complexity of multiple galactic components or time evolution. %
    In \autoref{sec:results:static_mw_lmc}, we consider a more complex \glsentrylong{MW}-like system perturbed by an \gls{LMC}-like satellite. 
    While the potential is still constructed from analytic components rather than a live simulation, it is considerably more complex than the triaxial NFW case, featuring strong non-axisymmetric structure and localized deviations from a smooth galactic background. %
    We also use this system to assess the Bayesian framework (PINN~\ref{model:pinnIV}-B and PINN~\ref{model:pinnV}-B), testing whether the framework yields good uncertainty estimates that track the regions of strong perturbation. %

    In \autoref{sec:results:evolving_mw_lmc}, we test the time-dependent extension by applying the PINN~\ref{model:pinnVI} variant to a dynamically evolving Milky Way-like system. Built from analytic components, this system features two sources of time evolution: a rotating galactic bar, and the orbit of the \gls{LMC}. %
    The key question is whether our time-dependent framework can learn a continuous evolution while 
    maintaining reconstruction accuracy comparable to the static case. %

    In \autoref{sec:results:fire}, we extend beyond analytic potentials by validating the framework on a realistic cosmological simulation drawn from the FIRE suite, where the gravitational potential exhibits complex structure that cannot be captured by analytic forms \citep{Wetzel2023}. %
    More specifically, we use a Milky Way-like galaxy from the \textit{Latte} suite of simulations~\citep{Wetzel2016} simulated with FIRE-2 physics~\citep{Hopkins2018}.
    We first evaluate the static model on a snapshot that approximates the present-day Milky Way (\autoref{sec:results:fire:static}). Next, we test the time-dependent model on a sequence of snapshots spanning $1.5 \ \mathrm{Gyr}$ of evolution, including the infall and orbit of an LMC-like satellite (\autoref{sec:results:fire:evolving}). %
    These simulations provide a stringent test of the framework in a setting where no ground truth potential is available, and hence the residual field cannot be described analytically. 

\begin{figure*}
    \centering
    \hspace{-1.25cm}
    \includegraphics[width=1.06\linewidth]{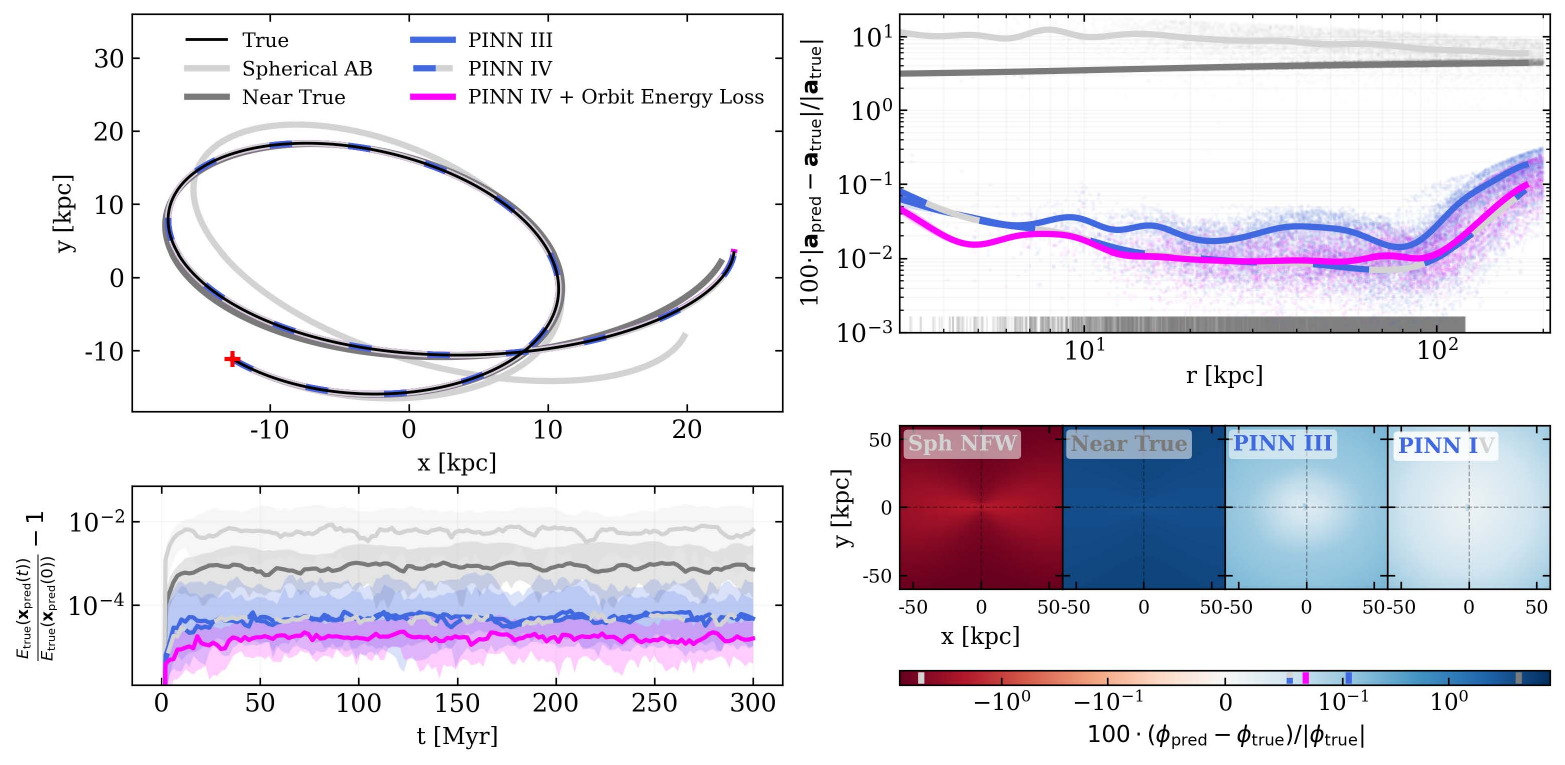}
    \caption{%
        \textbf{Triaxial NFW test system. }%
        Comparison of two \gls{PINN} variants vs two analytic benchmark models: %
        (i) a spherical NFW potential with the same mass and scale radius as the true halo (``Spherical AB''), and %
        (ii) a near-truth triaxial NFW potential with one axis ratio offset by $5\%$ relative to the true system (``Near True''). %
        \textit{Left: Dynamical diagnostics.} %
        A representative orbit integrated for $300\,\mathrm{Myr}$ in the true potential,  compared to the trajectories for the two learned models and the two analytic baselines. %
        The bottom panel shows the energy drift of predicted trajectories evaluated in the true potential, aggregated over a sample of 50 orbits. %
        The solid line marks the median deviation over all orbit samples, and the shaded band spans the minimum to maximum. 
        \textit{Right: Field-level reconstruction.} %
        The top panel shows radially binned acceleration errors. The colored lines denote the PINN variants, and the gray lines mark the two analytic benchmarks. The gray dashes on the horizontal axis mark the radii of the training points. %
        The four bottom panels show the relative potential residuals in the $x$--$y$ plane, with the median error for each marked by the colored line on the horizontal bar below. %
    }
    \label{fig:results:triaxial_nfw:triaxial}
\end{figure*}

    \subsection{Triaxial NFW} \label{sec:results:triaxial_nfw}

       As our first test case, we adopt a static triaxial NFW halo \citep{NFW:1997} as the ground truth potential. %
       A triaxial halo is a natural starting point since it represents a simple but realistic departure from spherical symmetry, with no additional substructure or time evolution to complicate the interpretation of the results. %
       We describe the system setup and training details below, before presenting the results in \autoref{fig:results:triaxial_nfw:triaxial}. %

        We denote the ground-truth potential by
        \begin{equation}
            \phi_{\mathrm{NFW}}^{\mathrm{triax}}(\mathbf{x}; M, r_s, q_1, q_2),
        \end{equation}
        which is implemented using the \texttt{galax} triaxial NFW potential \citep{galax}. %
        We use mass scale \(M = 10^{12}\,M_\odot\), scale radius \(r_s = 10\,\mathrm{kpc}\), and axis ratios \(q_1 = \tfrac{4}{5}\) and \(q_2 = \tfrac{5}{4}\). %


        We train several PINN models on position--acceleration pairs sampled from this true potential. %
        To contextualize their performance, we compare against two analytic benchmarks: %
        \begin{enumerate}[leftmargin=1.0em]
            \item A spherical NFW halo $\phi_{\mathrm{NFW}}^{\mathrm{sph}}(\mathbf{x}; M, r_s)$, representing a reasonable situation where the large-scale mass distribution is known, but the true triaxiality is not. %
            \item A near-truth triaxial NFW $\phi_{\mathrm{NFW}}^{\mathrm{triax}}(\mathbf{x}; M, r_s,  q_1', q_2)$ with $q_1' = 0.95\,q_1$ (a $5\%$ perturbation from the true axis ratio) representing a best-case analytic approximation where the model family is right but the parameters are slightly wrong. %
        \end{enumerate}

        The top left panel of \autoref{fig:results:triaxial_nfw:triaxial} shows orbits integrated in the true potential (black) and the two analytic benchmarks from the same initial conditions. %
        Note that the orbit in the spherical potential (light gray) strongly departs from the true trajectory, consistent with the large pointwise acceleration errors of $\sim 10\%$ across the entire domain (top right). %
        The near-truth triaxial potential (dark gray) remains more faithful to the true trajectory but diverges near the end of the integration, and also has large acceleration errors of $\sim5\%$ across the domain. %
        Both analytic models perform poorly in potential reconstruction, with average reconstruction errors in the $x$--$y$ plane exceeding 5\% (bottom right). %
        The failure of even well-motivated analytic models to reproduce the true dynamics motivates the use of the PINN framework, which can absorb these discrepancies through a learned correction. %

        To compare PINN performance with these analytic benchmarks, we train two PINN variants which differ only in whether an analytic baseline is included. %
        The first is PINN~\ref{model:pinnIII} (\autoref{fig:results:triaxial_nfw:triaxial}, solid blue line), where the network predicts a scaled proxy for the full potential. %
        We treat the dominant radial dependence as an NFW profile, and factor it out of the network output using the scaling function %

            \begin{equation}\label{eq:scale-factor}
            n(\mathbf{x}; r_s)
            \;=\;
            \frac{r}
            {r_s\ln\!\left(1 + \dfrac{r}{r_s}\right)},
            \qquad
            r = \|\mathbf{x}\| .
        \end{equation}

        %

        so that the potential is reconstructed as described in \autoref{eq:phi_scaled}. %
        The second model is PINN~\ref{model:pinnIV} (\autoref{fig:results:triaxial_nfw:triaxial}, blue+gray line), which augments the scaled residual with an analytic baseline. %
        Here, we choose the baseline to be the spherical benchmark model, i.e. an NFW halo with the same $(M,r_s)$ as the true potential but no triaxiality, %
        \begin{equation}
            \phi_{\mathrm{AB}}(r)
            \;=\;
            \phi_{\mathrm{NFW}}^{\mathrm{sph}}(\mathbf{x}; M, r_s).
        \end{equation}
        The full potential is then reconstructed as described in \autoref{eq:phi_fused}. %
        In this variant, the neural network is tasked only with learning the triaxial residual relative to the well-motivated spherical baseline. %

        For the training datasets, we draw $4{,}096$ samples from a spherical volume ($r < 120 \ \mathrm{kpc}$) via density-based rejection sampling: each point is accepted with probability proportional to the local mass density of the true potential. %
        This concentrates the training data in the inner regions where the potential has the most structure, while still providing coverage at larger radii. %
        Each position is paired with the exact acceleration computed from the ground-truth analytic potential. %
        
        We train all models for $4{,}000$ epochs under an exponentially decaying learning rate (see \autoref{sec:appendix:training_details} for specific training details). %
        To assess the orbit energy-conservation term described in \autoref{sec:methods:orbit_energy}, we train an additional PINN~\ref{model:pinnIV} variant (\autoref{fig:results:triaxial_nfw:triaxial}, pink line) using the expanded loss function in \autoref{eq:combined_loss}, which includes both the acceleration-based loss and the orbit energy loss. %
        For this variant, we augment the acceleration training set with $50$ orbits pre-computed in the true potential. 
        These orbits are initialized from positions drawn uniformly in log-radius between $1\,\mathrm{kpc}$ and $120\,\mathrm{kpc}$, and integrated for $500\,\mathrm{Myr}$ in the true potential, with $50$ equally spaced snapshots saved per trajectory. %
        To prevent the orbit-based loss from dominating the optimization, we set $\lambda_{\mathrm{E}}^{\max} = 0.2$ and during training use the ramped weighting formulation as described in \autoref{sec:methods:orbit_energy}. 

        We choose the training length of $4{,}000$ epochs to balance high accuracy and low computational cost, with the training for each PINN model converging in $< 4$ minutes on an Apple M1 CPU. %
        Since the additional design features (i.e. inclusion of an analytic baseline) represent transformations of the network output rather than additional learned parameters, they add minimal computational cost. %
        Beyond $4{,}000$ epochs, additional training does not yield a meaningful improvement in performance for any variant.

        \autoref{fig:results:triaxial_nfw:triaxial} compares the performance of all models against the ground truth potential. 
        We assess performance along two axes: pointwise acceleration accuracy and long-term dynamical consistency. %
        Starting with pointwise accuracy, both PINN variants significantly outperform the analytic reference models, including the near-truth triaxial NFW (\autoref{fig:results:triaxial_nfw:triaxial}, dark gray). %
        PINN~\ref{model:pinnIII} achieves a maximum acceleration error of $0.3 \%$ in the extrapolation regime, with errors below $0.1 \%$ across the entire training domain  (top right). %
        The inclusion of analytic fusion in PINN~\ref{model:pinnIV} yields improved pointwise acceleration accuracy across all radii, with errors below $0.03 \%$ across the majority of the domain, reaching maximum of $0.25 \%$ only in extrapolation. %
        This improvement highlights that learning residual structure relative to a physically motivated baseline is more effective than learning the full potential directly. %
        Similar trends carry over to the potential reconstruction (\autoref{fig:results:triaxial_nfw:triaxial}, bottom right), with PINN~\ref{model:pinnIV} achieving a mean relative error of $0.06\%$ that improves upon the $0.11\%$ error reached by PINN~\ref{model:pinnIII}. %
        Notably, the energy conservation term also acts as a global regularizer: the variant trained with this additional 
        loss (``PINN IV + Orbit Energy Loss", pink) further reduces acceleration errors in the inner region ($r < 20\,\mathrm{kpc}$) relative to PINN~\ref{model:pinnIV}, suggesting that the orbit-based constraint provides complementary information beyond what the pointwise acceleration loss alone captures. %

        Turning to the dynamical performance, we compare orbits integrated in the learned potentials with the true trajectory (\autoref{fig:results:triaxial_nfw:triaxial}, top left). %
        All three PINN models are visibly indistinguishable from the true trajectory, confirming that the pointwise accuracy translates into long-term dynamical consistency. %
        We also evaluate energy conservation along each trajectory: for all variants, the energy drift within the learned potential is negligible (relative deviations $\sim10^{-7}$). %
        This is expected: because the model predicts a scalar potential and derives accelerations through $\mathbf{a} = -\nabla\phi$, the resulting dynamics are inherently conservative, so energy is preserved up to numerical integration error. %
        The small observed drift therefore reflects integrator accuracy rather than a property of the learned field. %
        
        To more meaningfully assess the energy consistency, we evaluate the energy of the predicted trajectories within the true potential, $E_{\mathrm{true}}(\mathbf{x}_{\mathrm{pred}})$, rather than within the learned potential. %
        This probes how well the learned dynamics reproduce the true Hamiltonian, rather than tracking internal consistency within the learned potential. %
        We emphasize that this quantity is used only for post-training evaluation; during training, the energy regularization term requires only orbit trajectories and does not assume knowledge of the true potential. %
        Both analytic benchmarks perform poorly under this metric, with the spherical and near-true triaxial potentials reaching relative errors of $\sim 10^{-2}$ and $\sim10^{-3}$ respectively, reflecting the mismatch between their trajectories and the true Hamiltonian (\autoref{fig:results:triaxial_nfw:triaxial}, bottom left). %
        PINN~\ref{model:pinnIII} and  PINN~\ref{model:pinnIV} improve substantially, achieving relative deviations of $< 10^{-4}$ across the entire integration. %
        Including the energy conservation loss provides a further $\sim 20\%$ improvement, confirming that orbit-based regularization improves not just internal consistency but genuine dynamical fidelity. %

        Overall, this experiment demonstrates that each design feature contributes meaningfully to reconstruction accuracy: spatial scaling improves numerical stability, and adding an analytic baseline further improves both local force accuracy and long-term orbit behavior. %
        Even when the analytic baseline is intentionally misspecified, PINN~\ref{model:pinnIV} outperforms a near-truth analytic approximation, highlighting that learned residuals can compensate for imperfect prior knowledge of the potential. %
        This robustness is critical for force-fitting and force-replay applications \citep[e.g.,][]{Arora_2024,Petersen:2022:EXPNbodyIntegration}, where flexible representations are needed to model realistic force fields.

        \begin{figure*}[t]
            \centering
            \hspace*{-1.4cm} \includegraphics[width=1.11\linewidth]{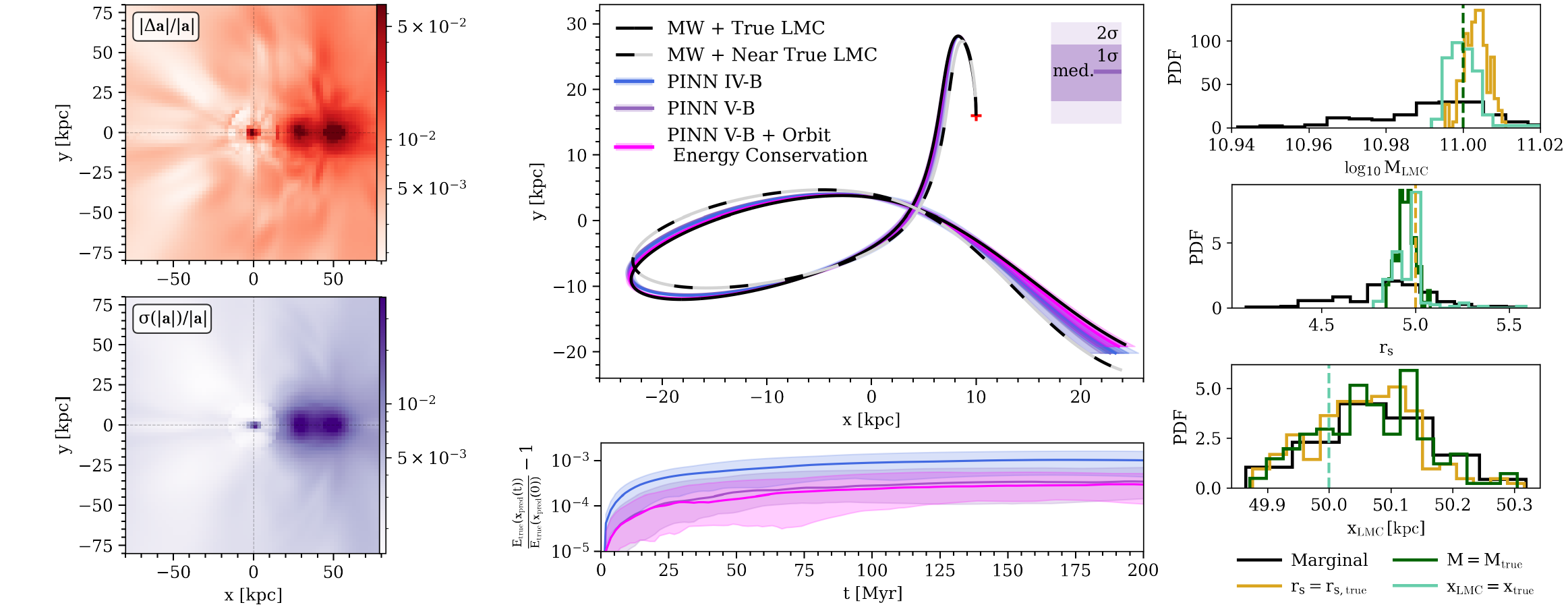}
            \caption{%
                \textbf{\gls{MW}--\gls{LMC} test system.} %
                \textit{Left: Field-level reconstruction.} %
                Median relative acceleration residuals (top) and relative posterior acceleration uncertainty (bottom)  from the most complex model (PINN~\ref{model:pinnV}-B + Orbit Energy Conservation) in the $x$--$y$ plane, aggregated over 150 posterior draws. %
                \textit{Center: Dynamical diagnostics.} %
                Test-particle orbit initialized at the \gls{LMC} center and integrated for $200\,\mrm{Myr}$ in several models: the true potential, a \gls{MW}--\gls{LMC} model with misspecified parameters (\gls{LMC} center offset by $10\,\mrm{kpc}$ and scale radius $r_s$ misestimated by $20\%$), and three \gls{PINN} variants (described in \autoref{sec:results:static_mw_lmc}). %
                Top: orbit traces, where the solid lines mark the median trajectories over all posterior draws and the shaded bands show the $1\sigma$ and $2\sigma$ spreads. %
                The red cross marks the initial position. %
                Bottom: fractional energy drift of the predicted trajectories evaluated in the true potential, with the solid lines showing the medians and shaded bands showing the $1\sigma$ posterior spreads. %
                \textit{Right: Parameter inference.} %
                Posterior constraints on three \gls{LMC} parameters -- mass (top), scale radius (middle), and Galactocentric distance (bottom) -- inferred by PINN~\ref{model:pinnV}-B + Orbit Energy Conservation. %
                For each posterior draw, the full learned potential is fit with an analytic \gls{MW}--\gls{LMC} model to recover the \gls{LMC} parameters, assuming the true \gls{MW} potential; the resulting distributions are shown as histograms with dashed lines marking the true values. %
                The 1D marginal posteriors are compared to distributions obtained by conditioning on the true values of the other parameters. %
            }
            \label{fig:accel_orbit_errs}
        \end{figure*}


    \subsection{Milky Way + LMC} \label{sec:results:static_mw_lmc}

        Having established that the framework accurately reconstructs a simple triaxial potential, we now consider a more demanding test system: an analytic \gls{MW} potential perturbed by an \gls{LMC}-like satellite. %
        This system features the complexity of multiple galactic components and the non-axisymmetric perturbation of the LMC, making it a much stronger test of the framework's ability to learn residual structure. %
        We also test the Bayesian framework, assessing whether the predicted uncertainty tracks the regions of strong perturbation. %
        We first describe the system setup before presenting the results in \autoref{fig:accel_orbit_errs}. %
        The true potential consists of the following components: %
        \begin{itemize}[leftmargin=*]
            \item \textbf{Halo}: a triaxial NFW halo with mass $5.4\times10^{11}\,M_\odot$, scale radius $r_s = 15.62\,\mathrm{kpc}$, and principal-axis ratios $q_1=\tfrac{4}{5}$ and $q_2=\tfrac{5}{4}$. %
            The mass and scale radius follow the \texttt{MilkyWayPotential} of
            \citet{Price-Whelan2017}, itself based on \citet{Bovy_2015}; we
            introduce triaxiality not present in the \texttt{MilkyWayPotential} model as N-body simulations generically predict
            non-spherical dark matter halos \citep[e.g.][]{Jing_2002, Allgood:2005eu}. 
            \item \textbf{Disk}: a \gls{MN} disk \citep{Miyamoto:1975:ThreeDimensionalModelsDistribution} with mass $6.8\times10^{10}\,M_\odot$ and scale lengths $a = 3.0\,\mathrm{kpc}$ and $b = 0.28\,\mathrm{kpc}$, following \citet{Bovy_2015}.
            \item \textbf{Bar}: a Long--Murali bar \citep{Long:1992:AnalyticalPotentialsBarred} with mass $1.0\times10^{10}\,M_\odot$ and axis lengths $a = 4.0\,\mathrm{kpc}$, $b = 1.0\,\mathrm{kpc}$, and $c = 1.5\,\mathrm{kpc}$, with mass and semi-major axis consistent with observational estimates of the Milky Way bar \citep{Portail:2017:DynamicalModellingGalactic, Wegg:2015:StructureMilkyWays}. %
            \item \textbf{\gls{LMC}}: an NFW potential with mass $10^{11}\,M_\odot$, scale radius $r_s = 5.0\,\mathrm{kpc}$, and Galactocentric center at $(50, 0, 0)\,\mathrm{kpc}$. The mass and distance are consistent with current estimates \citep{2013ApJ...764..161K, Erkal+:2019}; the scale radius is chosen to give a reasonable inner mass profile for an NFW model of this total mass.
        \end{itemize}

        In the center of \autoref{fig:accel_orbit_errs}, we show an orbit integrated in the true potential (black). %
        For comparison, we also show an orbit integrated in a near-true analytic model (black+gray): this model uses the correct \gls{MW} but places the \gls{LMC} $20\%$ closer to the Galactic center and increases its scale radius by $20\%$. %
        Despite having the correct functional form and only mildly misspecified parameters, the trajectory visibly diverges from the truth, highlighting how even very accurate analytic models can fail to reproduce the correct orbital dynamics. %
        
        Motivated by the limitations of this analytic benchmark, we now test whether the PINN framework can succeed even when starting from a baseline that is missing major components. %
        As in \autoref{sec:results:triaxial_nfw}, we evaluate two variants that differ in how the analytic baseline is treated; both are Bayesian here, allowing us to assess uncertainty calibration alongside reconstruction accuracy. %
        The first, PINN~\ref{model:pinnIV}-B, uses a fixed analytic baseline consisting of a spherical NFW halo and a \gls{MN} disk --- deliberately missing both the bar and the \gls{LMC}. %
        The included components are also misspecified: the halo mass and scale radius are overestimated by $10\%$, while all disk parameters (mass and scale lengths $a$, $b$) are underestimated by $10\%$, mimicking the kind of systematic errors that arise when fitting a simplified mass model to observational data. %
        This is a deliberately challenging baseline, where major components are missing and the included components have incorrect parameter values, designed to test the limits of what the residual network can absorb. %
        
        The second, PINN~\ref{model:pinnV}-B, uses the same analytic functional form as PINN~\ref{model:pinnIV}-B but treats all five baseline parameters (halo and disk masses, halo $r_s$, and disk scale lengths) as free to vary during training. %
        By optimizing these jointly with the neural residual, the model can correct large-scale misspecification while simultaneously learning higher-order corrections without any prior knowledge of the correct parameter values. %
        For all variants, we use a spherical NFW scaling function for $n(\mathbf{x})$ with scale radius $r_s = 15.62\,\mathrm{kpc}$, matching the true halo scale radius and the non-dimensionalization scale. %

        We construct the training set by drawing $4{,}096$ position samples from a spherical volume via density-based rejection sampling (as described in \autoref{sec:results:triaxial_nfw}). %
        We sample data only from radii below $150 \ \mathrm{kpc}$, but test the model over a wider range ($r < 250 \ \mathrm{kpc}$). %
        We also train a third variant, {PINN~\ref{model:pinnV}-B + Orbit Energy Conservation}, which augments the training objective with the orbit energy conservation loss introduced in \autoref{sec:methods:orbit_energy}. %
        This variant uses the same acceleration training set supplemented with a set of $20$ reference orbits. %
        Their initial positions are drawn from the training set, with each initial velocity set to the local circular velocity. %
        Each orbit is integrated for $500\,\mathrm{Myr}$ in the true potential, with $50$ equally spaced snapshots saved per trajectory. %

        We train all PINN variants for $20{,}000$ epochs. %
        For PINN~\ref{model:pinnIV}-B, we use a single training stage with the fixed analytic baseline described above. %
        For PINN~\ref{model:pinnV}-B, we use the two-stage training schedule described in \autoref{sec:appendix:training_details}: an initial analytic-focused phase of $2{,}000$ iterations fits the baseline parameters to the large-scale structure, followed by a residual-focused phase of $18{,}000$ epochs that allows the network to capture higher-order corrections. %
        For PINN~\ref{model:pinnV}-B + Energy Conservation, we use the same two-stage schedule but introduce the orbit energy loss (\autoref{eq:energy_loss}) during the final $8{,}000$ epochs, ramping its contribution gradually via a cosine schedule to maintain optimization stability. %
        The total training time for each Bayesian model is approximately $12$ minutes, as measured on an Apple M1 CPU. %
        For comparison, deterministic training requires approximately $4$ minutes to reach comparable accuracy. %
        The additional computational cost reflects the need to explore the posterior distribution under stochastic variational inference (SVI). %
        We emphasize that all training was performed on CPU; the \GalactoPINNS{} code is fully GPU-compatible, and GPU acceleration substantially reduces training times. %

        We show the performance of all PINN variants in \autoref{fig:accel_orbit_errs}. %
        As shown in the top left, residual acceleration errors remain small across most of the domain (with a mean relative error of $0.62\%$) and peak only in the immediate vicinity of the \gls{LMC}, where the true potential departs most strongly from the analytic baseline. %
        The Bayesian posterior uncertainties closely track the spatial structure of these residuals (\autoref{fig:accel_orbit_errs}, bottom left), confirming that the predicted uncertainty is well-calibrated: largest where the reconstruction errors are largest, and smallest where the model is most reliable. %

        To assess whether this local improvement translates into globally consistent dynamics, we integrate test-particle orbits in the reconstructed potentials and compare them to ground-truth trajectories (\autoref{fig:accel_orbit_errs}, center). %
        As a robust test, we choose an orbit initialized at the center of the LMC, with an initial velocity set to the local circular velocity of the true potential at that position. %
        A key advantage of the Bayesian framework is that we can integrate orbits across multiple posterior draws, propagating the pointwise acceleration uncertainties directly into uncertainties on the trajectory; this is reflected in the shaded bands shown in the center panel of \autoref{fig:accel_orbit_errs}. %
        Even with a fixed, incorrect baseline, the median path learned by PINN~\ref{model:pinnIV}-B (blue) reproduces the true trajectory more accurately than the near-truth analytic model. %
        This highlights that learning a residual field can compensate for moderate analytic misspecification. %
        Allowing the analytic parameters to be optimized jointly with the residual in PINN~\ref{model:pinnV}-B (purple) further improves the orbit reconstruction: the median predicted trajectory remains within $2\sigma$ of the true trajectory at all points, despite minor deviation near the end of the integration. %
        Addition of the energy loss term (pink) yields the strongest performance: the learned trajectory shows no significant deviation from the true path across the entire integration, where the median deviation remains below $0.1\,\mathrm{kpc}$, compared to deviations of up to $6.1\,\mathrm{kpc}$ for the near-truth analytic model. %
        This improvement is consistent with the energy-loss interpretation introduced in \autoref{sec:results:triaxial_nfw}: penalizing energy drift along reference orbits suppresses spatially correlated force errors that the acceleration loss alone may miss. %


        As in \autoref{sec:results:triaxial_nfw}, we evaluate dynamical consistency by computing the energy of the predicted trajectories in the true potential. 
        Even without the additional energy loss term, the learned potentials already exhibit good energy stability along the integrated trajectories, with typical fractional deviations of $\sim 3 \times 10^{-4}$ for PINN~\ref{model:pinnV}-B (see \autoref{fig:accel_orbit_errs}, bottom center). Including the energy loss term provides a minor improvement in conservation, yielding deviations of $\sim 2 \times 10^{-4}$. The largest gains occur early in the trajectory, with continued improvement over the other PINN models across all 200 Myr of integration. 

        Beyond reconstruction accuracy and orbit fidelity, the Bayesian framework opens up an additional capability: the ability to extract physically interpretable constraints on the properties of the perturbing satellite directly from the learned potential. %
        Such constraints are relevant both for understanding the LMC’s dynamical impact on the Milky Way \citep{Erkal+:2019, Shipp_2021, Petersen_2020, Erkal_2021, Cunningham_2020, Magnus+Vasiliev:2021, Garavito-Camargo_2024} and for interpreting satellite-induced perturbations in cosmological simulations \citep{Wetzel2023, wetzel2025secondpublicdatarelease}. %
        In the right panels of \autoref{fig:accel_orbit_errs}, we extract posterior constraints on the \gls{LMC} mass, scale radius, and Galactocentric distance inferred by PINN~\ref{model:pinnV}-B + Orbit Energy Conservation. %
        We use $50$ samples drawn from the variational posterior to reconstruct the full learned potential (analytic plus residual); for each of these samples, we perform a least-squares fit over \gls{LMC} parameters while holding the \gls{MW} parameters fixed to their ground-truth values. %
        This probes how well the learned residual captures the \gls{LMC}-induced perturbation. %
        Accounting for parameter degeneracies by individually conditioning on true values, we recover all three \gls{LMC} parameters to within $2\sigma$. %
        The Galactocentric distance is especially well constrained, with a posterior mean of $50.04 \ \mathrm{kpc}$ that accommodates the true value ($50 \ \mathrm{kpc})$ to within $0.1\sigma$. %
        After marginalizing over the degeneracy with the LMC position, the LMC scale radius is recovered with a mean of $4.85 \ \mathrm{kpc}$, which includes the ground truth value ($5 \ \mathrm{kpc}$) to within $1 \sigma$. %
        We caution that these constraints assume perfect \gls{MW} reconstruction; in realistic settings where no ground-truth potential is available, parameter recovery would be correspondingly degraded. %

        Overall, this experiment demonstrates that the PINN framework remains robust even under major misspecification of the analytic baseline --- recovering accurate force fields and consistent orbits despite missing major components. %
        The results motivate extending the framework to time-dependent systems, where the potential evolves continuously due to satellite infall and bar rotation. %

        \begin{figure*}
            \centering
            \hspace{-1.15cm}
            \includegraphics[width=1.06\linewidth]{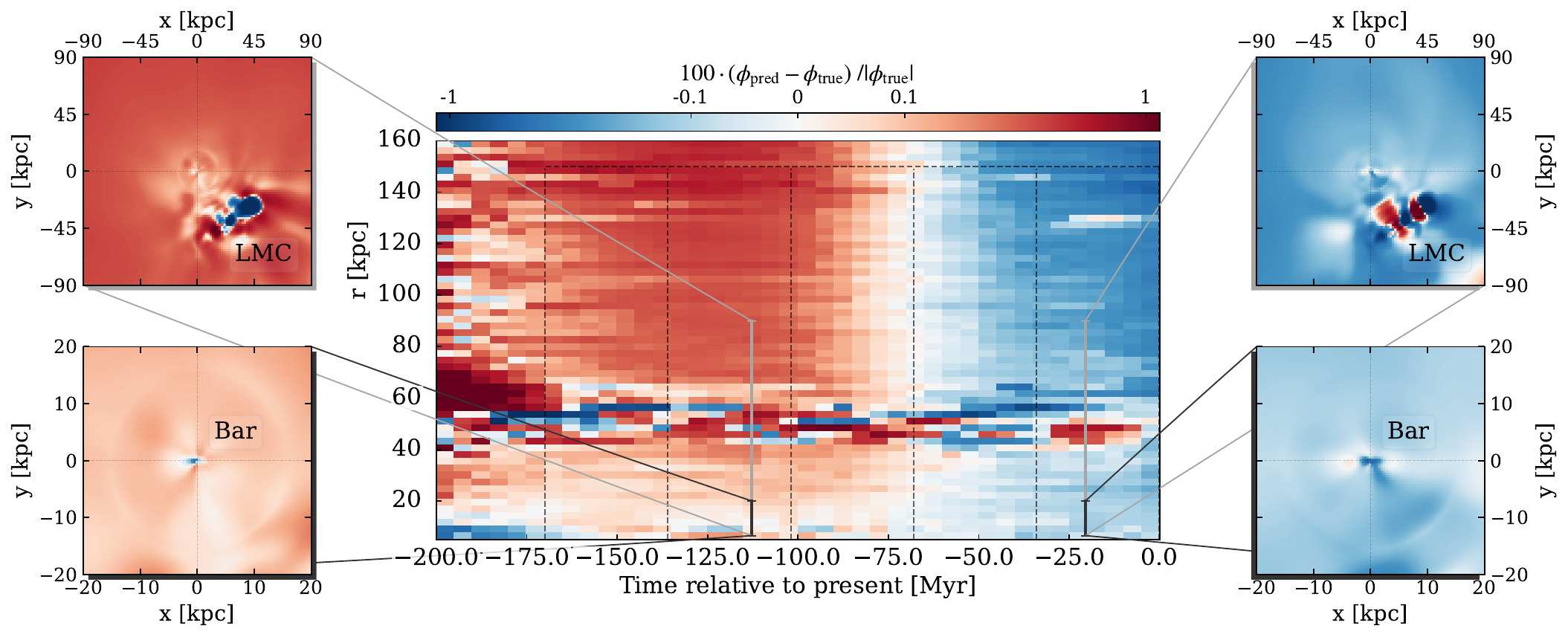}
            \caption{
            \textbf{Time-dependent \gls{MW}--\gls{LMC} potential reconstruction.} %
            \textit{Center:} Radially averaged relative error of the reconstructed \gls{MW}--\gls{LMC} potential, binned by radius and time relative to the present day ($t=0$). %
            The color scale shows $100 \cdot (\phi_{\mathrm{pred}}-\phi_{\mathrm{true}})/|\phi_{\mathrm{true}}|$ after gauge-fixing, where both $\phi_{\mathrm{pred}}$ and $\phi_{\mathrm{true}}$ are evaluated relative to their values at the far-field reference point $(x,y,z)_{\mathrm{ref}}=(0,500,0)\,\mathrm{kpc}$. %
            Dashed vertical lines indicate the training snapshot times, while the dashed horizontal line marks the maximum radius included in the training set ($150\,\mathrm{kpc}$). %
            \textit{Cut-outs:} Spatial slices of the relative potential error in the $x$--$y$ plane at two representative epochs: $t=-111\,\mathrm{Myr}$ (left) and $t=-22\,\mathrm{Myr}$ (right), shown at both large scales (top) and small scales (bottom). %
        }
            \label{fig:mwlmc_heatmap}
        \end{figure*}
        
    \subsection{Evolving Milky Way + LMC} \label{sec:results:evolving_mw_lmc}

        The previous two sections established that the framework can accurately reconstruct static potentials, even in the presence of strong, localized  perturbations. %
        We now extend the evaluation to time-dependent potentials, applying the neural integration formulation  (PINN~\ref{model:pinnVI}, \autoref{sec:methods:modeling_time_dependence}) to a dynamically evolving Milky Way-like system. %
        This system has two independent sources of temporal evolution: the orbital motion of the \gls{LMC} and the rotation of the bar. %
        Therefore, it represents a controlled but nontrivial setting in which the gravitational field evolves on different spatial and temporal scales, offering a strong test of our time-dependent framework. %
        We first describe the setup of the true potential and the details of the training set, before presenting the results in \autoref{fig:mwlmc_heatmap}. %

        The full time-dependent potential consists of a Milky Way background with two distinct sources of time evolution. Compared to the static case in \autoref{sec:results:static_mw_lmc}, we also include a central bulge to build toward a more complete and realistic Milky Way mass model. %
        The system consists of the following components: 
        \begin{itemize}[leftmargin=*]
            \item A triaxial NFW halo and an MN disk with the same parameters described in \autoref{sec:results:static_mw_lmc}, together with a 
            central bulge modeled by a Hernquist potential \citep{hernquist_potential} of mass $5.0 \times 10^{9}\,M_\odot$ and scale radius $1.0\,\mathrm{kpc}$, 
            following the \texttt{MilkyWayPotential} defaults of \citet{Price-Whelan2017}.
            \item A rotating Long--Murali bar with the same mass and scale parameters as in \autoref{sec:results:static_mw_lmc}, but rotating at a constant angular rate of $0.1\,\mathrm{rad\,Myr}^{-1}$ (${\approx}102\,\mathrm{km\,s^{-1}\,kpc^{-1}}$). 
            This is substantially faster than current observational estimates of the Milky Way bar pattern speed \citep[${\approx}35$–$40\,\mathrm{km\,s^{-1}\,kpc^{-1}}$;][]{Portail:2017:DynamicalModellingGalactic, Sanders_2019}, making the temporal reconstruction problem intentionally more challenging. %
            \item An orbiting \gls{LMC} modeled as a spherical NFW potential with mass $10^{11}\,M_\odot$ \citep{Erkal+:2019} and scale radius $r_s = 5.0\,\mathrm{kpc}$. %
            This relatively concentrated halo produces a stronger, more localized perturbation, providing a more demanding reconstruction test while remaining astrophysically plausible. %
        \end{itemize}

        We generate the time evolution of the system by specifying the present-day phase-space configuration of the \glsentrylong{MW} and the \gls{LMC} and integrating the combined gravitational field backwards in time. %
        The \gls{LMC} is initialized at its present-day Galactocentric position with a three-dimensional velocity $\mathbf{v}_{\mathrm{LMC}} = (-56,\,-219,\,186)\,\mathrm{km\,s^{-1}}$, corresponding to a characteristic velocity scale of $130\,\mathrm{km\,s^{-1}}$ \citep{van_der_Marel_2002}. %
        We evolve the coupled \gls{MW}--\gls{LMC} system backward from the present day on a uniform temporal grid, producing a smoothly interpolated, time-dependent field. %
        We regularize short-range interactions during the integration by imposing a minimum Coulomb impact parameter of $b_{\mathrm{min}} = 1\,\mathrm{kpc}$ \citep{Vasiliev_2020}. %
        
        We construct the training dataset from six temporal snapshots evenly spaced between the present day ($t=0\,\mathrm{Myr}$) and $170\,\mathrm{Myr}$ in the past ($t=-170\,\mathrm{Myr}$), capturing the recent dynamical evolution of the system. %
        For each snapshot, we draw $2{,}048$ positions within a spherical volume of $150 \ \mathrm{kpc}$, accepting each point by mass-density weighted rejection sampling.  %
        As for the previous two static systems, these positions are paired with the exact accelerations computed in the ground truth \gls{MW}--\gls{LMC} potential. %

        Rather than comparing variants as we do in the previous sections (\autoref{sec:results:triaxial_nfw} and \autoref{sec:results:static_mw_lmc}), we focus on evaluating one time dependent model, namely PINN~\ref{model:pinnVI} with a fixed analytic baseline. %
        This baseline consists of only the time-dependent \glsentrylong{MW} background (halo, disk, and bulge), but deliberately excludes both the rotating bar and the \gls{LMC}. %
        These omitted components must therefore be captured entirely by the learned residual field. %
        We also employ spherical NFW scaling for $n(\mathbf{x})$ based on the true scale radius of the halo ($r_s = 15.62 \ \mathrm{kpc}$). %
        
        We train the network for $12{,}000$ epochs using the neural integration formulation described in \autoref{sec:methods:modeling_time_dependence}, with full training details provided in \autoref{sec:appendix:training_details}. %
        The architecture comprises two multi-layer perceptrons (MLPs): the first predicts the initial spatial correction $\tilde{\phi}_{\mathrm{NN}_0}(\mathbf{x})$ (depth 6) and the second models the time derivative $\dot{\tilde{\phi}}_{\mathrm{NN}}$ (depth 3), corresponding directly to the two components of \autoref{eq:node}; both networks have a hidden width of 64. %
        We find that increasing the network width does not meaningfully improve performance, but making it deeper does at the cost of longer training time; see \autoref{sec:appendix:eval_details} for more discussion about these architecture choices. %
        We train all temporal snapshots jointly; sequential training slightly degrades performance, consistent with the need to learn a globally coherent temporal evolution. %
        
        We now turn to the assessment of the learned potential,  beginning with the radially binned error shown in the central panel of \autoref{fig:mwlmc_heatmap}, which provides a global view of how well the model reproduces the true potential across both radius and time. %
        Average potential errors remain below $2\%$ out to $150\,\mathrm{kpc}$ across the full $170\,\mathrm{Myr}$ training window. %
        The most significant errors occur near $50\,\mathrm{kpc}$ -- where the LMC trajectory is concentrated -- at times outside the training window.  %
        Temporal interpolation performs substantially better than extrapolation: within the training interval, errors remain generally stable and below $\sim 0.6\%$. However, we note that the error near the LMC tends to peak\textit{ between} training snapshots; for instance, the error between $0\,\mathrm{Myr}$  and $-34\,\mathrm{Myr}$ (both included in the training set) is largest at intermediate times, where the model has the least direct supervision.
        Outside the range of the final training snapshot, the radially-averaged error increases approximately linearly with time at all spatial scales, reaching $> 4\%$ near the LMC after $t = -200\,\mathrm{Myr}$ (far outside the training bounds). %
        This behavior is a direct consequence of the neural ODE formulation, which learns smooth evolution within the sampled window but is not explicitly constrained to reproduce long-term dynamics beyond it. %

        We pause to clarify one feature of the heatmap (\autoref{fig:mwlmc_heatmap}, center) that is not physical in origin, namely the local reduction in the radially averaged error near $t\sim -70\,\mathrm{Myr}$. %
        This arises from the choice of gauge used to evaluate the potential error. %
        In evaluation, we fix the gauge by specifying a reference point at which the potential is held constant when computing $\Delta\phi$. 
        Changing this reference point shifts the apparent location of the low-error region in the heatmap, indicating that the feature is not intrinsic to the learned dynamics but rather a consequence of how the potential difference is normalized. %

        While the radially binned error provides a useful global summary, it can mask localized features that are confined to specific regions of the potential. %
        We therefore also examine planar cross sections of the relative potential error, shown in the right and left panels of \autoref{fig:mwlmc_heatmap}. %
        We provide snapshots in the $x$--$y$ plane at two characteristic times: $t= -111\,\mathrm{Myr}$ (left) and $t= -22\,\mathrm{Myr}$ (right). %
        In these planar cross sections, the largest deviations are centered near the rotating bar and the \gls{LMC}, where the true potential departs most strongly from the analytic baseline and exhibits the most complex structure. %
        Even in these regions, however, the error does not exceed $3\%$, demonstrating that the model successfully captures localized non-axisymmetric perturbations. %
        Importantly, this is comparable to the peak error of the static reconstruction (\autoref{fig:accel_orbit_errs}), confirming that our time-dependent framework can achieve per-snapshot accuracy on par with its static counterpart. %
        The radially averaged errors in the central heatmap are systematically lower than those in the planar snapshots: radial averaging is dominated by the weakly perturbed outer potential, which dilutes the localized high-error regions near the \gls{LMC}. %

        A faint ring is also visible in the small-scale planar slices (\autoref{fig:mwlmc_heatmap}, top left and right) at $r_0 \sim 16 \,\mathrm{kpc}$, matching the non-dimensionalization scale $r_s = 15.62 \,\mathrm{kpc}$. %
        Although subdominant relative to other error sources, this artifact is a numerical imprint of the input transformation. %
        We discuss its origin and implications for the choice of scaling in \autoref{sec:discussion:rep_considerations:input_coords}. %

        The planar slices also show a structure of alternating positive and negative residuals, more pronounced at earlier times (\autoref{fig:mwlmc_heatmap}, top left). %
        These features remain localized near regions of complex structure (such as the LMC) and are subdominant relative to the overall reconstruction accuracy. We discuss the origin of this pattern, and its relationship to other potential representations, in \autoref{sec:discussion:time_dependence}. %
        
        Overall, this experiment demonstrates that the neural integration formulation can successfully learn smooth, temporally coherent corrections to a baseline potential, capturing both secular evolution and localized perturbations. This enables continuous interpolation between discrete simulation snapshots, providing a differentiable force field for downstream dynamical analyses and complementing basis-expansion approaches \citep{Arora_2024, Petersen:2022:EXPNbodyIntegration}.


    \subsection{FIRE Simulations} \label{sec:results:fire}

        In each of the previous test systems, we generated the residual field as the difference between two analytic potentials: one representing the ground truth, and another representing a misspecified guess. %
        We now aim to evaluate the framework in a more realistic setting where the residuals cannot be described analytically. To do so, we apply the framework to cosmological simulations from the FIRE suite \citep{Wetzel2023}, which model realistic galaxy formation,  including complex baryonic processes like star formation, stellar feedback, and gas dynamics. 

        We focus on the \texttt{m12b} simulation, first introduced in~\cite{GarrisonKimmel2019}, a \glsentrylong{MW}-like system hosting multiple satellite galaxies, including one comparable in mass and size to the \gls{LMC}. %
        Among the seven \glsentrylong{MW}-mass galaxies in the original Latte suite~\citep{Wetzel2016}, \texttt{m12b} is the closest analog to the observed \gls{MW}--\gls{LMC} system \citep{Garavito-Camargo_2024}, making it a particularly relevant test case for the methods developed in this paper. %
        In \autoref{sec:results:fire:static}, we apply the framework to a single snapshot of the \texttt{m12b} simulation which approximates the present-day Milky Way. In \autoref{sec:results:fire:evolving}, we extend the evaluation to our time-dependent model, learning the time-evolving field from multiple snapshots spanning $1.5 \ \mathrm{Gyr}$. %

        \begin{figure*}
            \hspace{-0.9cm}\includegraphics[width=1.06\linewidth]
            {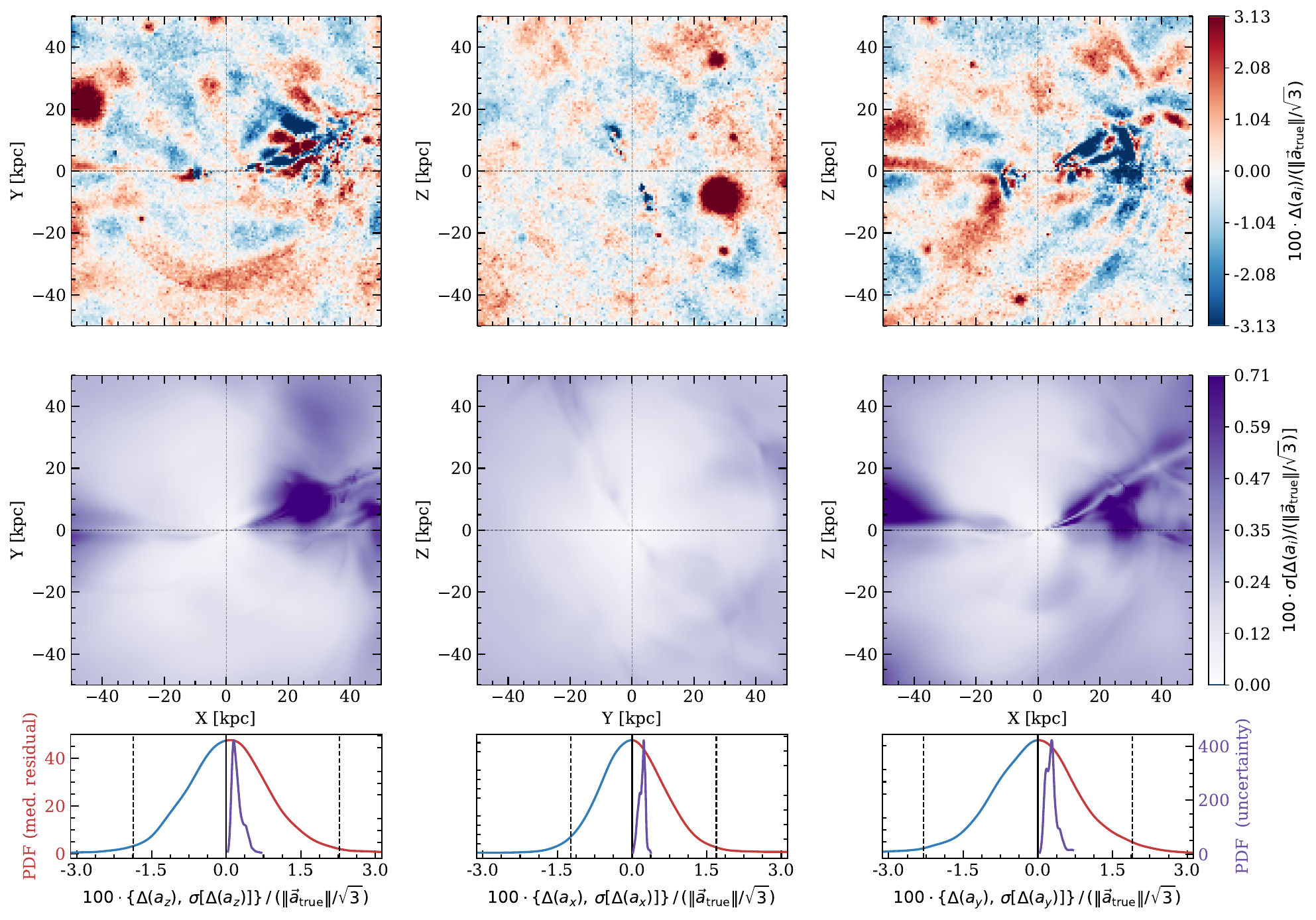}
            \caption{%
                \textbf{Static reconstruction of the \texttt{m12b} potential under Bayesian inference.} %
                \textit{Top row:} Median signed relative acceleration residuals in the $x$--$y$, $y$--$z$, and $x$--$z$ planes within a $50 \times 50\,\mathrm{kpc}$ region centered on the galaxy, computed over 500 posterior draws. %
                The plotted component is the acceleration component perpendicular to each slice: $a_z$ in the $x$--$y$ plane, $a_x$ in the $y$--$z$ plane, and $a_y$ in the $x$--$z$ plane. %
                The coordinate system is oriented so that the $x$--$y$ plane is aligned with the galactic disk. %
                \textit{Middle row:} Posterior uncertainties for the same acceleration components over the same spatial region and posterior samples. %
                Here $\Delta a_i \equiv a_{i,\mathrm{pred}}-a_{i,\mathrm{true}}$ denotes the signed residual in component $i$, and both residuals and uncertainties are normalized by the characteristic component scale $|\mathbf{a}_{\mathrm{true}}|/\sqrt{3}$. %
                \textit{Bottom row:} Distributions of the signed residuals shown in the top row, with negative residuals in blue and positive residuals in red, together with the uncertainty distributions from the middle row in purple. %
                Histograms are binned using Scott's rule, and dashed black lines mark the corresponding $2\sigma$ intervals. %
                Across the plotted region, the reconstructed forces have component-wise residuals below $4\%$ and posterior uncertainties below $1\%$. %
            }
            \label{fig:m12b_heatmap}
        \end{figure*}

        \subsubsection{Static field} \label{sec:results:fire:static}

            We first apply the PINN framework to a static snapshot of the \texttt{m12b} simulation, focusing on the epoch that most closely resembles the present-day Milky Way. %
            Following \citet{Garavito-Camargo_2024}, we select the snapshot corresponding to the first pericentric passage of the LMC-like satellite, which occurs $5.01\,\mathrm{Gyr}$ before the end of the simulation. %
            In this snapshot, the gravitational field reflects the full complexity of a realistic galaxy: complex baryonic and dark matter structure, plus the strong perturbation of a massive satellite -- beyond what was represented in our previous analytic models. %
            We describe the training setup below before presenting the results in \autoref{fig:m12b_heatmap}. %

            We construct the training set by sampling $16{,}384$ positions in the target gravitational field. %
            We choose a larger number of training points than the controlled MW--LMC system (from \autoref{sec:results:static_mw_lmc}) to reflect the greater complexity of the simulated field. Because the training points are drawn directly from the simulated distribution of particles, rejection sampling is not necessary: by randomly subsampling the source positions, the training and testing sets naturally reflect the structure of the field. %
            Unlike the analytic test systems, the true potential has no closed-form expression, so accelerations cannot be computed exactly. Instead, we evaluate the acceleration at each training position using the \texttt{Pytreegrav} library \citep{pytreegrav}, which computes gravitational forces via direct summation over all source particles. %

            This simulation provides a setting where 
            uncertainty quantification is especially valuable: the true potential cannot be fully described by any analytic form, meaning the model's predictions will carry larger errors in some regions, and knowing \textit{where} the reconstruction is least reliable is as important as the reconstruction itself. We 
            therefore adopt a Bayesian model (PINN~\ref{model:pinnV}-B), which provides 
            calibrated uncertainty estimates alongside the predictions.
            Since the true potential has no analytic form, there are no ground-truth parameters to recover. Therefore, we adopt a PINN~\ref{model:pinnV}-B model in which the analytic baseline is optimized jointly with the learned residual. %
            This baseline comprises a Hernquist bulge, \gls{MN} disk, and a spherical NFW halo, with all associated parameters (masses and scale lengths) treated as trainable. %
            To learn these parameters under Bayesian inference, we place truncated-normal priors on all analytic parameters, centered on values obtained from a least-squares fit to the training accelerations --- giving the 
            model a physically reasonable starting point. We set the prior ranges to $\pm 20\%$ around the best-fit values, reflecting the expected level of uncertainty from fitting a simplified mass model to observational data.
            This is designed to mirror a realistic application of our method, in which one can pre-fit an analytic potential and then use the network to soak up the remaining residual.
            Allowing additional low-impact components (e.g., those primarily affecting the nuclear region) to vary did not improve performance, though the framework can in principle accommodate a broader parameter set. %

            During training, we adopt a staged inference strategy similar to that in \autoref{sec:results:static_mw_lmc}, jointly optimizing the analytic and residual components via SVI for $92{,}000$ iterations. %
            We train in two main stages: a shorter analytic-focused phase ($8{,}000$ iterations), followed by a longer residual-focused phase ($84{,}000$ iterations) which learns the residual on top of the analytic model optimized in the previous stage. The prior widths on the residual weights are narrow in the first stage and widened in the second to enforce this ordering; full details are provided in \autoref{sec:appendix:training_details}. %
            We additionally apply importance weighting during training: each training point is weighted by the local residual between the true acceleration and the initial analytic baseline, so the network focuses its capacity on regions where the analytic model is least accurate 
            (\autoref{sec:appendix:training_details}).
        
            Since this system is more complex than the controlled test cases, we adopt a deeper network (depth 8) and a larger number of epochs. Combined with these choices, longer training is required for Bayesian models compared to deterministic training, since more time is needed to explore the posterior. Hence, training converges in $\sim 340$  minutes  on an Apple M1 CPU --- much longer than the previous test systems.
            For comparison, we also train a deterministic version of the same model, which converges in $\sim 20$ minutes. 
            We emphasize that for practical applications where uncertainty estimates are not required, the deterministic model's training time is more representative. %


            We present the performance of the Bayesian model in \autoref{fig:m12b_heatmap}, where the top row shows the component-wise acceleration errors, each evaluated in the plane transverse to that component's direction. In each plane, the errors remain below $\sim4.0\%$, with the largest deviations occurring in the $x$--$y$ plane. The coordinate system is defined such that the $x$--$y$ plane aligns with the principal axes of the disk, where the gravitational field exhibits the strongest non-axisymmetric structure and hence the model has the most difficult target. %
            The errors are systematically lower in the $x$--$z$ and $y$--$z$ planes, reaching 99th percentile errors of $3.2 \%$ and $3.1\%$, respectively (\autoref{fig:m12b_heatmap}, top). %
            In all planar snapshots, the errors are largest near $r \sim 40\,\mathrm{kpc}$, corresponding to the pericenter distance of the LMC-like satellite ($38\,\mathrm{kpc}$, \cite{Garavito-Camargo_2024}). 
            As in the analytic MW--LMC system (\autoref{fig:mwlmc_heatmap}), all three planar slices show a faint ring-like imprint at $r \sim 35\,\mathrm{kpc}$, the same non-dimensionalization artifact discussed in \autoref{sec:discussion:rep_considerations:input_coords}. %
            
            The predicted uncertainties, shown in the center row of \autoref{fig:m12b_heatmap}, generally track the error profiles, confirming that the posterior is well-calibrated: regions of higher reconstruction error correspond to regions of higher predicted uncertainty. %
            Indeed, the highest uncertainty is in the $z-$component of the acceleration as measured in the $x$--$y$ plane (center left), where the uncertainty saturates at a maximum of $0.8 \%$ in the same regions where the error is largest. 
            For all planar slices ($x$--$y$, $x$--$z$, and $y$--$z$), the maximum uncertainty is achieved near the LMC ($r\sim 40 \ \mathrm{kpc}$), with uncertainty plateauing outside this region. %
            
            For comparison, a deterministic version of the same model achieves a lower peak error of $\sim2.5\%$ in the $x$--$y$ plane, at the cost of providing no uncertainty estimates. %
            We discuss the trade-off between Bayesian uncertainty quantification and reconstruction accuracy in \autoref{sec:discussion:interpretability}. %
            
            Overall, this experiment demonstrates that the framework generalizes effectively to a realistic cosmological simulation, achieving sub-$4\%$ acceleration errors even in the presence of complex, non-analytic structure—and that the Bayesian extension provides good uncertainty estimates that reliably identify the regions where the reconstruction is most challenging. %
            These results demonstrate the feasibility of constructing accurate, uncertainty-aware force-field surrogates for cosmological simulations. %

        \subsubsection{Evolving field} \label{sec:results:fire:evolving}

           Having established that the framework accurately reconstructs the static \texttt{m12b} potential, we now test whether it can track the evolution of the same system over time --- a considerably harder problem. %
           Rather than learning a single snapshot, the model must recover a smoothly evolving field from a handful of discrete snapshots spread across several gigayears of cosmological evolution. %
            
           We use the extended FIRE release, which provides simulation outputs across $10\,\mathrm{Gyr}$ of evolution. 
           In this section, we define the time coordinate $t$ relative to the snapshot at which the LMC analogue reaches pericenter. %
           This snapshot is assigned $t=0$ and corresponds to a lookback time of $5.01\,\mathrm{Gyr}$ \citep{Garavito-Camargo_2024}. %
           With this convention, negative values of $t$ denote snapshots before pericenter, while positive values denote snapshots after pericenter.
           The training set comprises ten snapshots in total: eight evenly spaced between $t = 413\,\mathrm{Myr}$ and $t = -311\,\mathrm{Myr}$, plus two additional snapshots extending to $t = -833\,\mathrm{Myr}$. This intentionally uneven cadence tests whether the model interpolates smoothly regardless of snapshot spacing.
            
           We choose the training window to include the full arc of the LMC's approach: at the earliest training snapshot the satellite is $\sim 170\,\mathrm{kpc}$ from the Galactic center, reaching pericenter at $\sim 40\,\mathrm{kpc}$ at $t = 0$, before drifting back to $\sim 100\,\mathrm{kpc}$ at the latest snapshot. Since \texttt{m12b} is a two-pericenter system \citep{Garavito-Camargo_2024}, the second pericenter falls outside this window and is not included in the training or validation set. From each snapshot, we randomly subsample $16{,}384$ position--acceleration pairs, matching the training set size of the static reconstruction in \autoref{sec:results:fire:static}.

            We build a PINN~\ref{model:pinnVI}  model, which learns the time derivative of the potential to reconstruct a continuous evolution. We include a fixed (non-trainable) analytic baseline. %
            This baseline consists of the same pre-fit model used for the static potential: an NFW halo, Miyamoto-Nagai disk, and Hernquist bulge, with numerical parameters determined by least squares fit to the training accelerations at $t=0$. Importantly, we do not include an LMC in the model or any time-dependence in the baseline. 
            In principle, a more complex or a time-evolving baseline (as for our controlled \gls{MW}--\gls{LMC} system) could also be used, but we choose a simple baseline containing only information in the present-day snapshot. 
        
            Similar to the controlled \gls{MW}--\gls{LMC} test system, we use a simultaneous training strategy where all time snapshots are trained at once, rather than being passed in sequentially. %
            The architecture again comprises two MLPs: the first predicts the initial spatial correction $\tilde{\phi}_{\mathrm{NN}_0}(\mathbf{x})$ (depth 6) and the second models the time derivative $\dot{\tilde{\phi}}_{\mathrm{NN}}(\mbf{x},t)$ (depth 3); both have a hidden width of 64. %
            Training proceeds for $24{,}000$ epochs, where we again choose a longer training schedule to match the greater complexity of the field. Training converges in approximately 30 minutes on an Apple M1 CPU. %
            Similar to the static model from \autoref{sec:results:fire:static}, we use importance weighting to focus training capacity on the points that deviate most strongly from the analytic baseline. 
            Further training details are provided in \autoref{sec:appendix:training_details}. %

\begin{figure*}
                \hspace{-30pt}
                \includegraphics[width=1.07\linewidth]
                {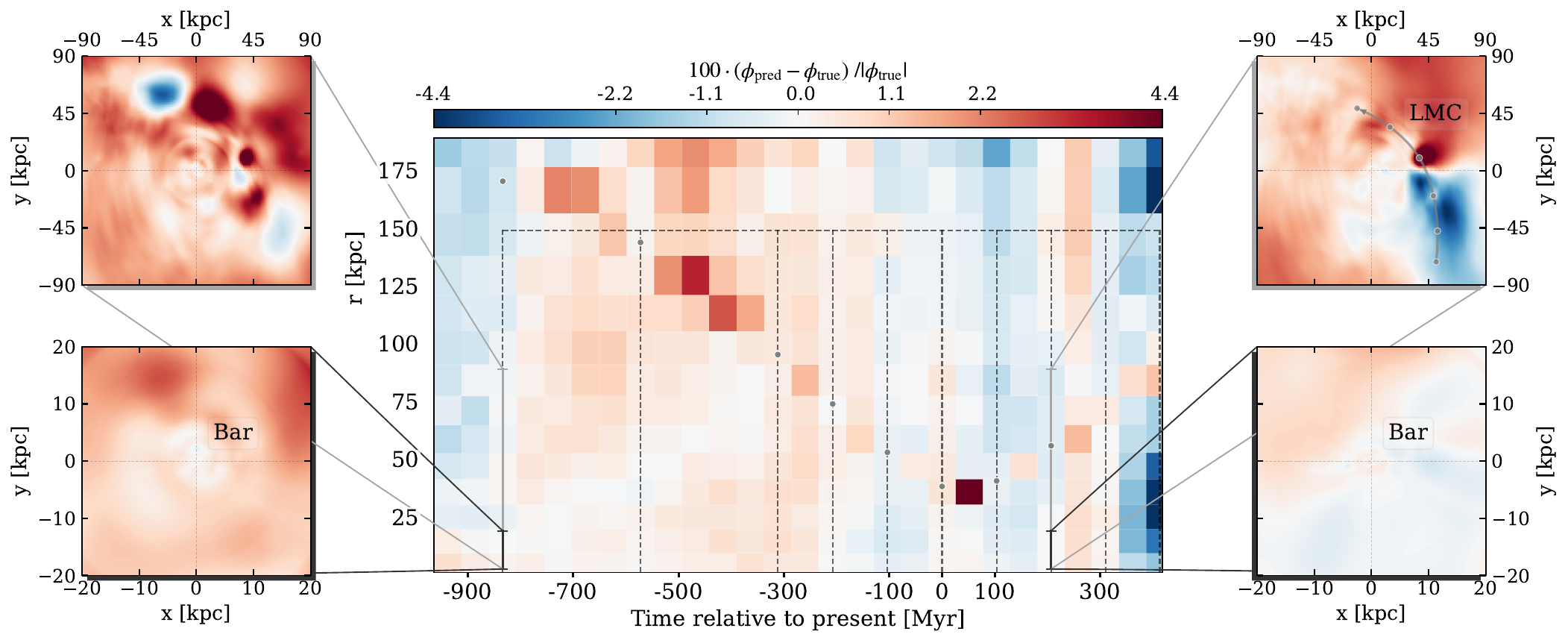}
                \caption{%
                \textbf{Reconstruction of the time-evolving \texttt{m12b} potential} across $1.5\,\mathrm{Gyr}$ of evolution. %
                \textit{Center:} Radially averaged relative error of the reconstructed \texttt{m12b} potential (after far-field gauge fixing),
                binned by radius and time relative to the present ($t=0$); negative times correspond to earlier epochs. %
                Dashed vertical lines indicate the training snapshot times, while the dashed horizontal line marks the maximum radius included in the training set ($150\,\mathrm{kpc}$). %
                The gray dots represent the position of the \gls{LMC} at each training snapshot. 
                Across the full $1.5\,\mathrm{Gyr}$ interval and out to the maximum training radius, the radially averaged errors remain below $5\%$. %
                \textit{Cut-outs:} Spatial slices of the relative potential error in the $x$--$y$ plane at two representative epochs: $t = -833\,\mathrm{Myr}$ (left) and $t = 206\,\mathrm{Myr}$ (right), shown at both large scales (top) and small scales (bottom). %
                }
                \label{fig:m12b_timedep}
            \end{figure*}

            Turning to an evaluation of the learned potential, we show the potential reconstruction error in \autoref{fig:m12b_timedep}. %
            Radially averaged errors remain below $5\%$ across the full evaluation window --- systematically higher than the controlled analytic test systems but reflective of the increased complexity of the field. %
            The largest errors occur along the trajectory of the LMC, which we plot in gray dots on the center heatmap. The largest radially averaged errors occur near $t =0$ and $r \sim 40 \ \mathrm{kpc}$ at the pericenter of the LMC, where the LMC is closest to the galactic center and interferes most strongly with the inner components. 
            Excluding the LMC, the reconstruction errors remain at the $\sim 1\%$ level across the full temporal window, showing the model accurately learns the smooth galactic background even with larger localized errors near the LMC. %
            Notably, temporal extrapolation performs substantially better here than in the controlled MW–LMC test system (\autoref{fig:mwlmc_heatmap}). %
            We discuss the origin of this difference, and the implications for choosing a training window, in \autoref{sec:discussion:time_dependence}

            In the left and right panels of \autoref{fig:m12b_timedep}, we show spatial cross sections of the 
            relative potential error in the $x$--$y$ plane at two characteristic times: $t = -833\,\mathrm{Myr}$ 
            and $t = 206\,\mathrm{Myr}$, chosen to contrast performance near the edge of the extrapolation regime with performance deep in the interpolation regime. %
            As in the controlled test systems, radially averaged errors are systematically lower than those visible in the planar snapshots, as radial averaging suppresses localized high-error regions and is dominated by the more weakly perturbed outer potential. %
            On large scales (top panels), the largest deviations are confined to the regions near the LMC ($r \sim 40\,\mathrm{kpc}$), where the strongest perturbation is located and the analytic baseline (which lacks an LMC) is most wrong. 

            The temporal resolution is constrained by the snapshot cadence of the \texttt{m12b} simulation. %
            Re-running the full cosmological simulation to generate additional outputs is computationally prohibitive, and therefore we use the existing snapshots for both training and validation. %
            Within this constraint, we construct the training set with uneven spacing, including both densely and sparsely spaced snapshots. %
            Notably, we find that performance in the sparsely sampled intervals is comparable to that in the densely sampled regime, indicating that uneven snapshot cadence does not significantly degrade reconstruction accuracy. %

            Overall, this experiment highlights that our time-dependent framework generalizes well beyond controlled analytic systems, achieving similar accuracy to the static reconstruction at each evaluation time. %
            The reconstruction remains accurate through the pericenter passage of an LMC-analog satellite, indicating that the framework can capture strongly time-dependent departures from equilibrium \citep{Wegg+:2019, Vasiliev_equilibrium}. %
            It also raises several considerations that arise when modeling cosmological simulations (rather than controlled analytic systems), which we discuss more in \autoref{sec:discussion:comparison}. %

\section{Discussion} \label{sec:discussion}

    The results presented in \autoref{sec:results} show that the \gls{PINN} framework accurately reconstructs galactic gravitational potentials across a wide range of physical complexity, from a controlled triaxial halo (\autoref{sec:results:triaxial_nfw}) to a 
    realistic cosmological simulation (\autoref{sec:results:fire}). %
    These gains accumulate progressively with each design feature rather than arising from any single architectural choice. %
    For static systems, our framework achieves sub-percent acceleration errors and stable orbit integration even when the assumed analytic baseline is missing major components such as the bar and \gls{LMC} (\autoref{sec:results:static_mw_lmc}). %
    For time-evolving systems, the neural integration formulation successfully captures smooth temporal evolution and localized non-axisymmetric perturbations, with radially averaged errors below $\sim\!4\%$ across all evaluation windows (\autoref{sec:results:evolving_mw_lmc}, \autoref{sec:results:fire}). %
    The Bayesian extension further enables uncertainty quantification and physically interpretable constraints on analytic baseline parameters. %

    \vspace{5pt}
    The remainder of this section examines the implications of these results in turn. %
    \autoref{sec:discussion:comparison} compares the PINN variants applied to two test systems, discussing which design features are most relevant to each application. %
    \autoref{sec:discussion:rep_considerations} examines  how representational choices -- such as the coordinate system, non-dimensionalization scales, and the smoothness of analytic components -- shape what the network must learn and how accurately it can do so. %
    \autoref{sec:discussion:interpretability} considers the role of Bayesian inference in interpreting and trusting the model's predictions. %
    \autoref{sec:discussion:relation_to_existing} situates the framework relative to existing potential modeling strategies. %
    \autoref{sec:discussion:time_dependence} addresses the implications of our time-dependent results for \gls{MW} modeling in the era of \textit{Gaia} and forthcoming surveys. %
    \autoref{sec:discussion:observational_considerations} discusses the challenges and opportunities in the application of our framework to observational data. %
    Finally, \autoref{sec:discussion:limitations} and \autoref{sec:discussion:future_directions} outline current limitations and promising directions for future work. %

    \begin{figure}
        \hspace{-1.1cm}\includegraphics[width=1.12\linewidth]{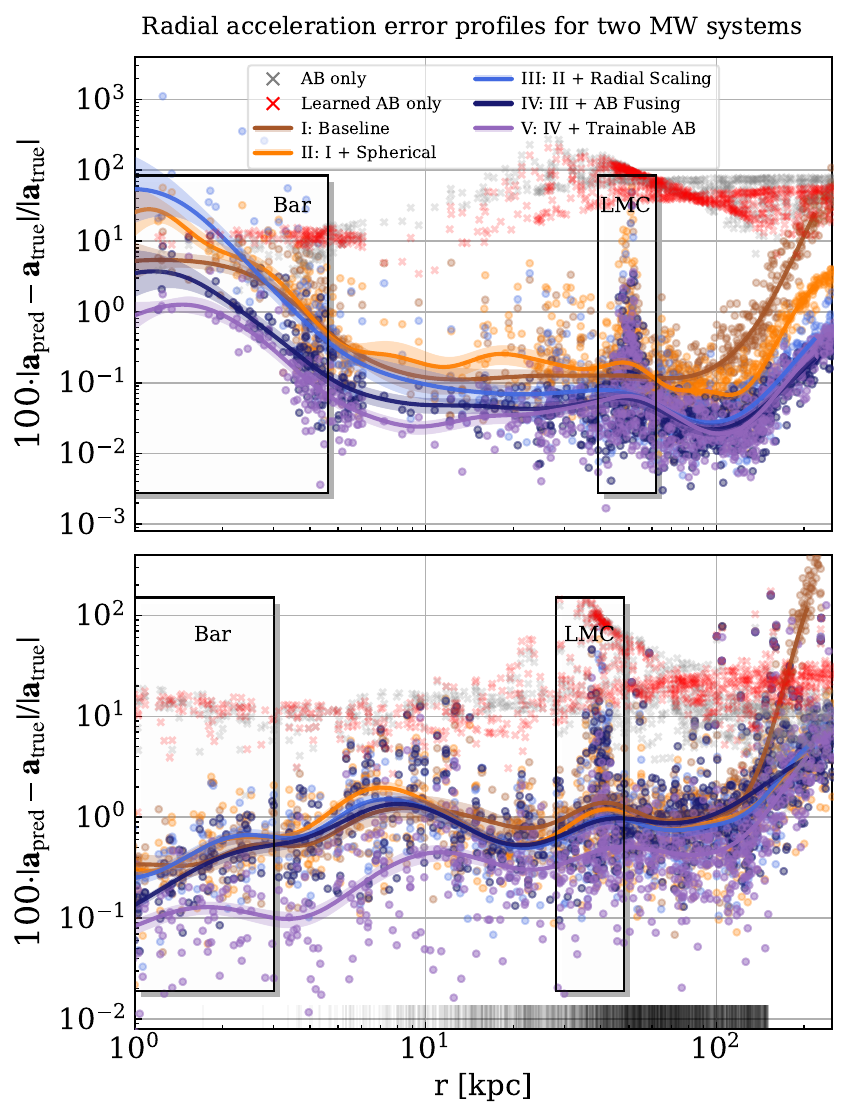}
        \caption{%
            \textbf{Cumulative impact of design choices on acceleration-field reconstruction.} %
            \textit{Top:} Radial profile of the relative acceleration error for the analytic \gls{MW}--\gls{LMC} test system. %
            We compare several \gls{PINN} variants (PINN~I--V); see \autoref{sec:methods:variants} for a description of the variants. %
            Points show pointwise acceleration errors at individual test locations, while solid curves show smoothed radial mean errors computed from logarithmic radius bins. %
            Black dashes along the horizontal axis mark the radii of the training points, and gray crosses show the error from an analytic \gls{MW} model without any learned correction. %
            The two raised black rectangles highlight oversampled test points in the most complex regions: near the Galactic bar on the left, and near the \gls{LMC} on the right, where the true potential deviates most strongly from the analytic baseline. %
            \textit{Bottom:} Same diagnostic for the static \texttt{m12b} simulation. %
        }

        \label{fig:ablation_mw}
    \end{figure}

    \subsection{Comparing PINN variants} \label{sec:discussion:comparison}

        The model variants introduced in \autoref{sec:methods:variants} are not all equally suited to every application; the appropriate choice depends on the complexity of the target system and what prior information is available. %
        To illustrate how the variants compare in practice, we evaluate PINN~\ref{model:pinnI} through PINN~\ref{model:pinnV} on two test systems: the analytic MW--LMC system from \autoref{sec:results:static_mw_lmc} and the cosmological \texttt{m12b} simulation from \autoref{sec:results:fire:static}. %
        This comparison also tests how the design choices translate across two different classes of potential --- one built from analytic components, and one drawn from a realistic cosmological simulation. %
    
        The top panel of \autoref{fig:ablation_mw} shows the performance of five PINN variants on the analytic MW--LMC system, where the training set is constructed as described in \autoref{sec:results:static_mw_lmc}. %
        All models are deterministic (since we focus on accuracy rather than uncertainty calibration) and are trained for $15{,}000$ epochs, with PINN~\ref{model:pinnV} using the two-stage schedule described in \autoref{sec:appendix:training_details}. %
        For comparison, we plot a benchmark analytic model (\autoref{fig:ablation_mw}, gray crosses): a spherical NFW halo plus Miyamoto--Nagai disk, omitting the LMC, the bar, and the triaxiality of the NFW halo. %
        This potential serves as the analytic baseline for PINN~\ref{model:pinnIV}, and exceeds acceleration errors of $100\%$ immediately around the LMC, where it differs most strongly from the true potential. %

        The simplest PINN model (PINN~\ref{model:pinnI}, brown) globally improves upon this, with average errors below $5\%$ across the entire training domain, only exceeding this in extrapolation. %
        The introduction of spherical coordinates (PINN~\ref{model:pinnII}, orange) brings significant improvement: at the edge of the training domain, the coordinate transformation reduces errors by a factor of $\sim 4$ (from $\sim 20\%$ to $\sim 5\%$). %
        In fact, without the coordinate transformation, PINN~\ref{model:pinnI} performs worse than the analytic benchmark in the far field, underscoring how sensitive the network is to input representation. %
        
        Spatial scaling by an NFW profile (PINN~\ref{model:pinnIII}, light blue) provides further improvement, particularly at large radii ($r \gtrsim 10\,\mathrm{kpc}$), where it systematically outperforms both simpler variants and the analytic benchmark. %
        Including an analytic baseline (PINN~\ref{model:pinnIV}, dark blue) improves performance globally, and allowing the baseline parameters to vary (PINN~\ref{model:pinnV}, purple) yields additional gains of $> 50\%$ in some regions, most notably near the bar where the fixed baseline is most misspecified. %
        Beyond $r \sim 50\,\mathrm{kpc}$, PINN~\ref{model:pinnIII}, \ref{model:pinnIV}, and \ref{model:pinnV} perform very similarly; the cumulative improvements are concentrated in the inner $20\,\mathrm{kpc}$, where the potential is most complex. %
    
        We also isolate the learned analytic component of PINN \ref{model:pinnV} (i.e. without the learned residual correction from the second stage of training), shown in red crosses in \autoref{fig:ablation_mw}. %
        The learned analytic model systematically improves on the fixed baseline (gray crosses), and in the far field ($r> 100\,\mathrm{kpc}$) it performs almost uniformly better. %
        This is not expected a priori, as we discuss further in \autoref{sec:discussion:interpretability}. %
    
        To probe performance in the most demanding regions, we evaluate all variants on test points sampled densely near the bar ($r < 5\,\mathrm{kpc}$) and the LMC ($r \sim 50\,\mathrm{kpc}$), shown as the raised boxes in \autoref{fig:ablation_mw}. %
        Near the bar, the most expressive variant (PINN~\ref{model:pinnV}) achieves a mean error below $1.5\%$ --- notable given that the analytic baseline includes no bar component, so the residual network must learn this feature from scratch. %
        The LMC remains the dominant source of error across all variants, producing the large spike at $r \sim 50\,\mathrm{kpc}$. %
        Even for PINN~\ref{model:pinnV}, the median error near the LMC stays below $0.1\%$, yet individual points can exceed $6\%$, a reminder that radially averaged metrics can mask localized failures and that sharp perturbations remain challenging to reconstruct. %
    
        We now repeat the ablation study on the \texttt{m12b} present-day snapshot, using the same training set described in \autoref{sec:results:fire:static}. %
        Each variant is trained for $15{,}000$ epochs, with PINN~\ref{model:pinnV} again following the two-stage strategy. %
        For the fixed-baseline variants, the analytic model now includes a Hernquist bulge alongside the \gls{MN} disk and spherical NFW halo, with all numerical values initialized via a least-squares fit to the training accelerations; the same fit sets the scale radius of the NFW spatial scaling function used by PINN~\ref{model:pinnIII} and subsequent variants. %
        The bottom panel of \autoref{fig:ablation_mw} summarizes the performance of all variants on this test system. %
        The pre-fit analytic benchmark (gray crosses) reaches maximum errors above $\sim 100\%$, highlighting the importance of residual learning for realistic potentials, while each design feature further improves performance and PINN~\ref{model:pinnV} achieves the lowest errors globally.
        
        The most significant gains between variants are concentrated in the inner region ($r < 10\,\mathrm{kpc}$) and in the extrapolation regime ($r > 150\,\mathrm{kpc}$). %
        The coordinate transformation (PINN~\ref{model:pinnII}) prevents the divergence at $r \sim 150\,\mathrm{kpc}$ seen in PINN~\ref{model:pinnI} by mapping radial coordinates to a bounded range (\autoref{sec:appendix:5d_coords}). %
        Spatial scaling (PINN~\ref{model:pinnIII}) yields modest improvement over PINN~\ref{model:pinnII} in the outer regions. %
        Further gains come from including an analytic baseline (PINN~\ref{model:pinnIV}), 
        which yields a factor of $\sim2$ improvement at $r < 3\,\mathrm{kpc}$ where the field is most complex; at larger radii, much of the structure is already captured by the spatial scaling and the gains plateau. %
        Making the baseline trainable (PINN~\ref{model:pinnV}) provides the greatest improvements, particularly in the inner $20\,\mathrm{kpc}$, where the median error remains consistently below $0.5 \%$. %
        As in the analytic MW--LMC system, the largest errors occur near the LMC: PINN~\ref{model:pinnV} achieves the lowest radially averaged error of $\sim 1\%$ in this region, though pointwise errors can still approach $\sim 5\%$.   %
    
        The inner-region errors are overall lower here than in the analytic MW--LMC system, likely because the analytic baseline includes a bar component, reducing the burden on the residual network in the most complex inner regions. %
        We also note that the difference between the learned and fixed analytic components of PINN~\ref{model:pinnV} is much more subtle than in the first test system: since the baseline is already initialized from a least-squares fit to the data, it provides the best possible analytic description from the outset, leaving less room for the trainable parameters to improve upon it. %
    
        In general, both test systems show that embedding more physical structure into the networks leads to improved performance, and that these gains are most effective when coupled with prior knowledge of the system (e.g., using an estimated or pre-fit scale radius for the NFW halo). %
        The two panels of \autoref{fig:ablation_mw} also show that analytic and realistic potentials require different modeling considerations and yield different performance, even when designed to simulate the same physical system. %
        We emphasize that the contribution of each design feature is not always fully apparent in isolation; some features provide modest gains on their own but become more consequential in combination with others, such as spatial scaling paired with analytic fusing. %
        However, the comparison of variants does support the general picture that performance scales with greater model complexity. %

    \subsection{Representational considerations} \label{sec:discussion:rep_considerations}

        The success of the framework depends not only on the complexity of the target potential, but also on a set of representational choices --- the coordinate system, the non-dimensionalization scales, and the analytic components --- that shape what the network must learn. %
        Here we discuss the implications of these choices, drawing on patterns from across the results. %

    \subsubsection{Input coordinates}\label{sec:discussion:rep_considerations:input_coords}

        Working in spherical coordinates is a natural choice for galactic potentials, which are generally organized around a central mass distribution \citep{BT08}. %
        This choice has a direct consequence on the structure of the reconstruction errors: because the network operates in spherical coordinates, errors tend to inherit a spherical character, appearing as roughly concentric features in the planar cross sections rather than Cartesian grid artifacts. %
        This effect is visible in \autoref{fig:accel_orbit_errs} (left),  \autoref{fig:m12b_heatmap} (top), and \autoref{fig:m12b_timedep} (left and right), where the error profiles follow the underlying spherical symmetry of the field. %
    
        Another consequence of the coordinate choice is the faint ring-like artifact visible in the planar error maps of \autoref{fig:mwlmc_heatmap}, \autoref{fig:m12b_heatmap}, and \autoref{fig:m12b_timedep}. %
        This radius corresponds to the NFW scale radius used for data non-dimensionalization (i.e. $r = 15.62 \ \mathrm{kpc}$ for \autoref{fig:mwlmc_heatmap}), which introduces a characteristic scale at which the representation of the potential transitions between inner and outer regimes; see \autoref{sec:appendix:scaling} for more details about this transformation. %
        Near this radius, the gradients of the scaled quantities change most rapidly, producing a small numerical imprint in the learned residual. %
        While this artifact is subdominant relative to other sources of error, it highlights the importance of choosing physically motivated scales for non-dimensionalization --- a poorly chosen scale radius could amplify this effect and degrade reconstruction accuracy near that radius.  %

    \subsubsection{Analytic components}\label{sec:discussion:rep_considerations:components}
    
        Although analytic potentials are parametrically simple and physically interpretable, they can introduce numerical challenges that are less pronounced in realistic systems. Commonly used forms such as the NFW profile have sharp central cusps, which generate large local gradients and increase stiffness in the learning problem. %
        Even with careful feature engineering and radial scaling, these sharp features can be difficult for neural networks to represent accurately, especially when combined with strong perturbations. %

        As a result, network performance depends not only on the complexity of the physical system, but also on the smoothness of the analytic components used to construct or approximate it. %
        In controlled experiments, we find that substituting smoother analytic profiles can significantly improve learning. %
        For example, replacing an NFW halo with a cored Burkert profile \citep{Burkert2000} improves stability near the origin and reduces reconstruction errors globally. %
        Similarly, adopting a Kuzmin disk \citep{Kuzmin_1956} yields improved performance in inner regions, as its simpler two-dimensional structure is easier for the network to represent than the more vertically extended Miyamoto--Nagai form. %
         
        While neither the Burkert profile nor the Kuzmin disk are physically motivated substitutes in the galactic context (the former lacks the NFW cusp characteristic of cold dark matter halos \citep{NFW:1997}, and the latter neglects the vertical structure of a realistic disk), these comparisons illustrate a general principle: profiles with sharp cusps or steep gradients place a greater burden on the residual network than smoother alternatives, regardless of how well they describe the true mass distribution. %
        Larger networks or longer training times can accommodate more complex profiles, but smoother analytic components lead to more efficient learning and better-conditioned optimization. %
        Although these choices are most immediately relevant for analytic test systems, the same principle applies when choosing analytic baselines for realistic potentials: smoother baselines can reduce the burden on the residual field and improve numerical conditioning. %

        The choice of analytic components also shapes how the potential behaves at large radii, with direct consequences for reconstruction. %
        Unlike the terrestrial systems studied by \citet{pinngm}, galactic potentials are not spatially confined; the mass distribution extends to arbitrarily large radii, and common profiles such as NFW have divergent total masses. %
        Consequently, imposing strict asymptotic boundary conditions (e.g., $\phi \rightarrow 0$ as $r \rightarrow \infty$) is not always physically or numerically appropriate. %
        
        In practice, the right choice depends on the nature of the target potential. For controlled analytic systems, we found that enforcing hard boundary conditions consistently degrades reconstruction accuracy, producing pronounced error spikes near the transition region. %
        In these cases, far-field behavior is more robustly captured through physically motivated radial scaling, which encodes asymptotic structure without introducing discontinuities. %
        For the \texttt{m12b} simulation, by contrast, a boundary transition has a negligible or mildly positive effect, since the simulated potential is smoother and lacks the sharp divergences of analytic profiles, making the boundary treatment better matched to the underlying field.

        For time-dependent potentials, the boundary treatment requires additional care. A fixed transition radius may become ill-posed if the potential evolves significantly at large radii (for instance, due to an infalling satellite whose influence extends to large distances at early times). We do not present results with a boundary transition in this paper, but it may still be appropriate in settings where the outer potential remains well described by a smooth analytic potential throughout the evaluation window.
        
        Taken together, these results emphasize that the choice of analytic components is not merely a modeling convenience but has a direct bearing on reconstruction quality.  %
        Well-chosen coordinate systems and analytic components reduce the burden on the residual network, making them important considerations when applying the PINN framework to new systems. %

    \subsection{Interpretability and uncertainty} \label{sec:discussion:interpretability}
    
        For realistic systems, the analytic baseline is rarely known with high precision, and the decomposition into ``baseline + residual'' is generally non-identifiable without additional constraints. %
        In practice, multiple combinations of analytic parameters and residual fields can produce similarly accurate reconstructions, implying that individual components should be interpreted probabilistically rather than as unique physical quantities. %
        
        This is especially relevant when interpreting the learned analytic parameters of PINN~\ref{model:pinnV} 
        (\autoref{fig:ablation_mw}, top panel). %
        Although the learned baseline outperforms the fixed one, this improvement is not guaranteed a priori: because the analytic and neural components are optimized jointly, their individual contributions are not uniquely separable, and the learned analytic parameters need not correspond to physically correct values. %
        In fact, the learned analytic model may in some cases be less accurate than a fiducial analytic choice, while the fused analytic+residual solution is more accurate overall. %
        The goal of the decomposition is therefore not strict parameter recovery, but rather a flexible representation in which global structure and local corrections can adapt together. %

        More broadly, reconstructing a potential from sparse local constraints is intrinsically ill-posed without strong priors or boundary conditions. %
        Spatial scaling and fusing with an analytic baseline are physics-informed priors, but their usefulness depends on whether the chosen baseline captures the main features of the system. %
        Common observational profiles (such as the Einasto model \citep{Einasto:1969}, which has been widely applied to describe dark matter halos) provide physically motivated starting points for such baseline choices. %
        When the baseline family is significantly misspecified, the residual network can absorb missing structure, but interpretability of the decomposition is reduced and uncertainty calibration becomes essential. %

        Under the Bayesian framework, treating both analytic parameters and network weights probabilistically provides two key advantages. %
        First, the spatially varying predictive uncertainty provides a principled indicator of where the reconstruction is least reliable. %
        As shown in \autoref{sec:results:static_mw_lmc} and \autoref{sec:results:fire:static}, the posterior uncertainty is consistent with the model being most uncertain where the true potential departs most strongly from the baseline. %
        Second, the posterior over analytic parameters provides not just point estimates but uncertainties on physically meaningful quantities such as halo mass and scale radius, giving a probabilistic picture of how well the large-scale mass model is constrained by the data. %
        
        At the same time, the variational approximation adopted here is fully factorized and therefore unable to capture correlations between analytic parameters and residual degrees of freedom.
        As a result, posterior samples should be interpreted as calibrated uncertainty indicators rather than definitive parameter constraints. Additionally, overall performance is modestly degraded relative to the deterministic model due to the mean-field assumptions of the variational approximation \citep{SVIhoffman2013stochasticvariationalinference, SVIwingate2013automatedvariationalinferenceprobabilistic}.
        When parameter inference itself is the primary objective, more expressive variational families (such as low- or full-ranked multivariate normal distributions \citep{Gelman_2013a}) or Monte Carlo-based approaches (such as the No-U-Turn Sampler (NUTS) of \cite{NUTShoffman}) may be required.


    \subsection{Relation to existing potential modeling strategies} \label{sec:discussion:relation_to_existing}

        It is useful to situate this approach relative to three broad classes of methods for modeling galactic potentials. %
        These approaches differ primarily in how they balance interpretability, flexibility, and computational efficiency. %

        Analytic models remain the most indispensable tool in galactic dynamics. %
        Parametric forms such as NFW halos provide fast and interpretable descriptions of galactic potentials and are implemented in many modern dynamical toolkits such as \texttt{galax} \citep{galax}. %
        However, these models are structurally limited: non-axisymmetry, satellite-induced perturbations, and time dependence must be introduced explicitly, and they often struggle when the true potential departs significantly from the assumed functional form. %
        The hybrid framework developed here is not intended to replace analytic potentials, but rather to extend them. %
        By treating an analytic model as a baseline and learning a residual field on top of it, the approach preserves the interpretability and efficiency of the analytic description while providing additional representational flexibility. %

        Basis function expansions (\glspl{BFE}) provide a second widely used strategy for increasing flexibility. %
        In this approach, the potential is represented as a sum over orthogonal basis functions with coefficients determined by fitting to the mass distribution or gravitational field \citep{BFEs}. %
        BFEs offer a systematic route to improving fidelity: increasing the number of modes allows progressively finer structure to be represented. %
        In practice, however, accurately capturing localized or highly non-axisymmetric features often requires a large number of terms, which can introduce numerical noise and reduce interpretability. %
        Low-order truncations are therefore commonly used, but they inevitably limit the fidelity of the reconstructed field. %

        The formulation developed here can be viewed as a generalization of BFEs, where the correction term is learned adaptively from data rather than fixed to a predetermined set of modes. %
        In this sense, the neural network provides a flexible, data-driven basis for the residual, one that can concentrate capacity where the potential departs most strongly from the analytic baseline, rather than distributing it uniformly across all spatial scales. %
        This prevents the need for unphysical ``counterweighting" terms (i.e. negative mass components) that are often needed to make BFEs globally consistent. %
        At the same time, the physics-informed loss constrains this representation, ruling out solutions that are accurate pointwise but physically inconsistent -- something that purely empirical basis expansions cannot guarantee. %

        More recently, machine-learning methods have been developed to infer gravitational potentials directly from phase-space data or stellar kinematics \citep[e.g.][]{Buckley:2023:MeasuringGalacticDark, Green:2023:DeepPotential, Kalda2023, KaldaGreen2025}. %
        These approaches typically define likelihood-based objectives in which the potential enters implicitly through orbit integration or equilibrium assumptions. %
        These methods are particularly well suited to observational inference problems where the potential must be inferred from tracer populations rather than from direct force measurements. %
        The approach presented here is complementary: instead of learning the potential implicitly from tracer statistics, we learn the gravitational field directly when acceleration information is available, such as in simulations or differentiable dynamical models. %

        Taken together, the proposed framework occupies an intermediate position between analytic modeling and fully data-driven inference. %
        It retains the interpretability and computational advantages of analytic potentials, extends their flexibility through a learned residual, and incorporates physical constraints through a physics-informed objective. %
        This combination makes the method well suited to systems where analytic models capture the dominant structure but fail to represent localized perturbations, non-axisymmetric features, or mild time dependence, and to both simulation force replay and empirical inference from acceleration data, where flexible yet interpretable potential models are required \citep{Arora_2024, Chakrabarti2020}. %


    \subsection{Time dependence and the MW--LMC context} \label{sec:discussion:time_dependence}
    
        The \gls{MW}--\gls{LMC} experiments presented here are not intended solely as stress tests of model flexibility, but instead reflect a broader shift in how the \glsentrylong{MW} potential must be modeled. %
        An ever-growing body of work shows that the \glsentrylong{MW} is not well described by a static, equilibrium potential: massive satellites such as the LMC induce coherent, time-dependent perturbations, non-inertial effects, and large-scale density wakes in both the stellar and dark-matter components \citep{Gomez2015, Cunningham_2020, Vasiliev_2020, Garavito-Camargo_2024}. %
        These perturbations produce measurable signatures in stellar kinematics, halo structure, and phase-space distributions \citep{Petersen_2020, Erkal_2020, Erkal_2021,  Pace_2022, ou2025decodinggalactictwirldownfall}. %
    
        At the precision enabled by \textit{Gaia} \citep{GaiaMission, GaiaDR3} and upcoming surveys, assumptions of a static or inertial frame become increasingly limited \citep{Vasiliev_2020}. %
        Even modest time-dependent distortions can bias dynamical inferences, especially when reconstructing global quantities such as halo shape, satellite masses, or orbital histories. %
        Accurate modeling therefore requires frameworks that can represent smooth temporal evolution while retaining enough flexibility to capture non-axisymmetric structure. %
        
        The neural integration formulation introduced in this work aims to address this need by representing the potential as a time-evolving field whose residual component is governed by a neural dynamical system. %
        This construction enforces temporal smoothness by design and provides a differentiable interpolation between discrete snapshots of the potential, avoiding the discontinuities that arise when treating each snapshot independently.
        
        While the model does not explicitly impose a physical evolution equation for the potential (and hence does not represent a true state-dependent ODE), the learned temporal dynamics remain constrained by the physics-informed loss and by consistency with the training trajectories.
        In this sense, the method occupies a pragmatic middle ground: it does not attempt to solve the full dynamical system, but instead learns a smooth and differentiable approximation that is sufficiently flexible to capture realistic perturbations.
        This makes it well suited for applications where only sparse temporal sampling is available, such as cosmological simulations or observationally inferred potentials.

        A key limitation of the neural integration formulation is its extrapolation behavior. As shown in \autoref{sec:results:evolving_mw_lmc} and \autoref{sec:results:fire:evolving}, errors grow approximately linearly outside the training window, though they remain at a level acceptable for most dynamical applications. For those requiring high-fidelity reconstruction at earlier times, we recommend extending the training window to include additional snapshots, as the model's accuracy degrades predictably and proportionally to the extrapolation interval.
        
        Another consequence of the integration-based representation is the introduction of small phase lags in the learned potential. These appear as a patchy structure in the spatial residual maps (see \autoref{fig:mwlmc_heatmap}, top left), where localized regions of positive and negative error alternate. This patchiness is slightly less pronounced in the FIRE reconstruction (see \autoref{fig:m12b_timedep}, top left), highlighting that the effect is less problematic for smoother, more realistic potentials. 
        This behavior suggests that the artifact is a product of small accumulated errors in the learned time derivative, which are amplified in systems with sharper analytic features.
        Notably, these residuals remain spatially localized around the LMC and the bar, rather than inducing compensating structure elsewhere in the domain. This contrasts with basis-function expansions discussed in \autoref{sec:discussion:relation_to_existing}, where localized perturbations are represented through globally supported modes and can therefore require oscillatory counterweights or ringing in distant regions of the galaxy.
        In this sense, the lumpy residual pattern seen here is better interpreted as a local phase error in the learned time evolution than as a global representational failure.
        In any case, more physically constrained dynamical formulations --- such as models that directly learn density evolution or explicitly enforce dynamical consistency ---  could mitigate these lag effects and improve long-term fidelity. 

        One natural extension of our framework is the incorporation of a time-evolving analytic baseline. In realistic settings, we often have prior knowledge about the dominant sources of temporal perturbation; for instance, the approximate orbit of the LMC can be constrained by proper motion measurements, and this information could be encoded directly into the baseline rather than left for the residual network to learn. Concretely, one could augment the analytic baseline with a time-dependent component, such as an NFW potential whose center follows a trainable spline trajectory, allowing the model to learn the satellite's orbit alongside the residual field. This would reduce the burden on the network to reconstruct the large-scale perturbation, while remaining flexible enough to accommodate observational uncertainties. By placing bounds on the spline knots (i.e. by constraining the satellite position to within a $10\ \mathrm{kpc} \times 10 \ \mathrm{kpc} \times 10 \ \mathrm{kpc}$ box in each snapshot), this approach could succeed even when the trajectory is only approximately known. 
        
        In preliminary experiments, we found that incorporating such a time-evolving baseline reduced radially averaged errors to below $2 \%$ near the LMC --- a meaningful improvement beyond the $\sim 4\%$ errors achieved in the time-evolving \texttt{m12b} reconstruction (\autoref{sec:results:fire:evolving}). To keep the method as simple and as general as possible, we choose not to adopt this approach, since it requires specifying the functional form and the parameterization of the time dependence in advance. However, for applications where strong perturbations are known and their approximate trajectories are available, adding a time-evolving baseline is a promising route to improved accuracy with minimal added complexity. 
        

    \subsection{Observational considerations}\label{sec:discussion:observational_considerations}

        The experiments in this paper are trained on acceleration samples drawn from known potentials (analytic models or simulations), but the framework is designed to ingest acceleration data of any origin. %
        This is increasingly relevant as direct, model-independent acceleration measurements in the Milky Way become available. %
        Precision pulsar timing now provides direct measurements of the Galactic acceleration field across kiloparsec scales \citep{Chakrabarti2020, Moran+:2024:PulsarbasedMapGalactic, Donlon+:2025:WeighingMilkyWays}, complemented by stellar acceleration constraints on the local potential \citep{PhysRevLett.123.091101, Silverwood_Easther_2019, PhysRevLett.127.241104} and Solar System acceleration measurements within the Milky Way \citep{Bovy:2020:PurelyAccelerationbasedMeasurement}. %
        These datasets remain sparse but demonstrate the feasibility of direct acceleration-based inference, and are expected to expand in both volume and spatial coverage with upcoming observations. %
        The framework is therefore naturally positioned to incorporate such heterogeneous acceleration datasets as they grow, without requiring equilibrium or distribution-function assumptions. %

        A practical challenge for any future acceleration-based inference is the sparse and inhomogeneous coverage of current datasets. Pulsar timing measurements, while precise, are confined primarily to the Solar neighborhood \citep{Moran+:2024:PulsarbasedMapGalactic, Donlon2024} and do not yet provide coverage of the outer halo or regions far from the Galactic plane -- precisely the regions where LMC-induced perturbations are strongest \citep{Cunningham_2020, Garavito-Camargo_2024}. 
        The Bayesian framework introduced here partially mitigates this by providing spatially varying uncertainty estimates that grow in regions of reduced coverage, flagging where the reconstruction should not be trusted. More directly, the analytic baseline provides a physically-motivated prior that anchors the reconstruction in data-sparse regions, reducing the risk of unphysical extrapolation. As acceleration catalogs grow in volume and spatial coverage, our framework can be straightforwardly extended to incorporate new measurements, with the analytic baseline continuing to anchor the reconstruction wherever direct constraints are absent. 
        


    \subsection{Limitations} \label{sec:discussion:limitations}

        Several limitations follow directly from the modeling choices. 
        Although a scalar potential supports density estimation via Poisson's equation, $\nabla^2\phi \propto \rho$, second derivatives of neural networks can be noisy and sensitive to both architecture and optimization \citep{wang2023expertsguidetrainingphysicsinformed}. %
        As a result, density reconstruction is not emphasized here, and the method should not currently be viewed as a drop-in replacement for approaches that infer mass density profiles directly. %
        Enforcing additional smoothness priors, using architectures designed for higher-order derivative stability, or incorporating weak supervision on density could improve this. %

        Although orbit fidelity improves empirically (see \autoref{fig:results:triaxial_nfw:triaxial} and \autoref{fig:accel_orbit_errs}), these gains arise indirectly rather than from any explicit enforcement of dynamical structure. The training objective guarantees only that the field is conservative, but does not constrain subtler dynamical invariants such as phase-space volume conservation (Liouville's theorem) or the existence of action-angle variables. For applications where long-term orbital accuracy is critical (i.e. modeling stellar streams) incorporating additional physics-based constraints, such as symplectic consistency losses or structure-preserving integrators, could further improve dynamical fidelity \citep{Wisdom+Holman:1991:Symplectic, 2004shdbookL, chen2020symplecticrecurrentneuralnetworks}.
        
        Our time-dependent model promotes temporal smoothness and causal consistency by learning the time derivative of the residual field and integrating it across time. %
        In the simulation-based experiments presented here, multiple snapshots spanning hundreds of megayears are available for training --- a luxury that real observational data do not afford. %
        In practice, observational constraints on a time-dependent potential must be inferred indirectly, for instance from the present-day kinematics of stellar populations that have been perturbed at different epochs \citep{Helmi:2018:MergerThatLed, BlandHawthorn:2016}, or from the morphology of tidal streams whose progenitors were disrupted at known times \citep{PriceWhelan:2016, Nibauer2022}. %
        Applying the neural integration formulation in such settings would require careful thought about how to define training snapshots from observational data, and how to handle the much sparser temporal sampling that observations provide. %
        
        Beyond the question of data availability, the method does not explicitly enforce physical evolution laws (e.g., continuity or Poisson-consistency constraints tied to an evolving density field), and its reliability therefore depends on the temporal coverage of the training snapshots and the inductive bias of the chosen parameterization.
        As a result, interpolation within the training interval is consistently accurate, while extrapolation beyond the temporal range of the data remains weakly constrained.
        Incorporating explicit evolution constraints or embedding additional physical symmetries into the network architecture may provide stronger temporal extrapolation.

    
    \subsection{Future directions}\label{sec:discussion:future_directions}

        The most direct application to observations would be to first test our method on sparser data, taking into account only line-of-sight acceleration measurements, which are the ones available in~\cite{Donlon2025LocalDMDensity}. Such a test will enable us to determine the expected error from our method based on data currently available.

        As a complementary track, we can also replace acceleration supervision with losses defined on stellar kinematics and tracer distributions. %
        This could be done by differentiably integrating orbits under the learned potential and matching summary statistics (dispersion profiles, likelihoods from distribution functions, etc.), thereby aligning the objective with what is measurable. %

        Additional physics-informed constraints could further improve stability and interpretability. These include enforcing Poisson consistency between the learned potential and the implied mass distribution, incorporating explicit boundary conditions at large radii, and adding orbit-based regularizers that enforce symplectic structure \citep{hamiltonNN, chen2020symplecticrecurrentneuralnetworks}. %
        Architectures designed for smooth derivatives (e.g., spectral representations or constrained parameterizations) could also reduce noise in higher derivatives  \citep{smoothNN, sitzmann2020implicitneuralrepresentationsperiodic, rackauckas2021universaldifferentialequationsscientific}. 
        More fundamentally, future approaches may benefit from architectures that encode orbit dynamical invariants directly, for example through Poisson normalizing flows that impose energy conservation at the structural level rather than introducing it as an additional loss term \citep{xu2022poissonflowgenerativemodels, souveton2025hamiltoniannormalizingflowskinetic}. %

        The Bayesian framework adopted here uses a fully factorized variational approximation, which is computationally efficient but cannot capture all correlations between parameters. %
        In practice, this means the posterior uncertainties should be interpreted as approximate indicators of model confidence rather than precise probability statements. More expressive approaches --- such as normalizing-flow posteriors or targeted MCMC sampling for a few analytic parameters --- could improve 
        the fidelity of uncertainty estimates and better separate the contributions of the analytic baseline and neural residual \citep{Gelman_2013a}. While these methods carry higher computational cost, they may be worthwhile when precise parameter inference is the primary goal, such as constraining the mass or position of a satellite from kinematic data.

        A complementary improvement would be to use the model's own uncertainty estimates to guide where new training points are sampled. Rather than sampling uniformly or by mass density, one could preferentially add points in regions where the model is most uncertain (i.e. near satellites or non-axisymmetric structures), concentrating the training budget where it is most needed and potentially achieving better accuracy with fewer data points overall.




\section{Conclusions} \label{sec:conclusions}

    Accurate models of galactic gravitational potentials are fundamental to understanding galaxy structure, formation history, and dark matter distribution, yet realistic potentials are neither static nor axisymmetric, and no single analytic form can capture their full complexity. %
    This motivates hybrid approaches that combine analytic structure with flexible learned corrections.
    This work frames galactic potential reconstruction as a hybrid inference problem in which analytic structure provides a physically motivated starting point, and a neural residual supplies the flexibility needed to represent higher-order structure and localized perturbations on top of this. %
    Across controlled analytic test systems and cosmological simulations, the central finding is that combining (i) a potential-based formulation with an acceleration loss, (ii) spherical input coordinates, (iii) radial scaling, and (iv) analytic fusing with (v) trainable parameters yields strong pointwise accuracy while improving dynamical behavior relative to purely analytic baselines. %
    The Bayesian treatment further enables uncertainty quantification that is qualitatively consistent with model error, and the neural integration formulation provides a practical constraint on temporal evolution beyond treating time as an unordered input coordinate. %

    A recurring theme across the results is that model capacity is most effective when it is reserved for departures from a known large-scale trend. %
    Radial scaling reduces the dynamic range of the target and prevents the optimization from wasting capacity on learning the dominant fall-off of the potential. %
    Analytic fusing then shifts the learning problem from approximating the full potential to approximating a residual field that is more localized and higher order. %
    In the triaxial NFW system, this manifests as a clear improvement over a model-misspecified spherical baseline and improved performance even against a near-true analytic model, indicating that the residual network can efficiently represent modest symmetry breaking without requiring a high-order basis expansion. %

    The same logic is more consequential in the \gls{MW}--\gls{LMC} setting, where the baseline intentionally omits and misspecifies components. %
    In that regime, our framework does not merely improve pointwise errors relative to benchmark analytic models; it also improves dynamical metrics such as orbit fidelity and energy conservation. %
    This is an important practical point: potential learning that is evaluated only via local acceleration residuals can appear accurate while still yielding orbit drift if the learned field violates global consistency constraints. %
    Although the present objective does not enforce Hamiltonian structure explicitly, the improved orbit behavior suggests that learning a scalar potential and differentiating it is already a strong regularizer compared to learning accelerations directly. %
    
    The addition of each sequential design feature improves the acceleration reconstruction. In reconstructing the fields of both the analytic and FIRE simulation Milky Way, the most expressive model variant (PINN~\ref{model:pinnV}), which includes a trainable analytic baseline, provides the best performance (see \autoref{fig:ablation_mw}). Crucially, this improvement carries over to the time-dependent domain: the neural integration model achieves potential reconstruction accuracy in individual time snapshots comparable to that of the static reconstructions, demonstrating that temporal coherence can be enforced without sacrificing pointwise fidelity (see \autoref{fig:mwlmc_heatmap} and \autoref{fig:m12b_timedep}). Even when major sources of perturbation (such as the bar and LMC) are entirely absent from the analytic baseline, the framework recovers the induced structure with sub-percent to low-percent errors, highlighting its robustness to baseline misspecification. 

    The main conclusions and recommendations of this work are as follows: 
    \begin{enumerate}
        \item Retaining physically-motivated analytic components in the model is both practically and interpretatively important: analytic components constrain the large-scale structure, reduce the burden on the residual network, and yield parameters with direct physical meaning --- making the model more reliable and easier to validate against existing knowledge of the system. 
        \item Allowing the PINN to learn the remaining residual field enables accurate force reconstruction even when the analytic baseline is missing major components or carries incorrect parameter values --- demonstrating that the framework is robust to the kind of model misspecification that is unavoidable in realistic applications. 
        
        %
        \item Embedding physical structure directly into the network architecture --- through modified spherical coordinates, spatial scaling, and fusing with an analytic baseline --- provides consistent and cumulative improvements in accuracy, with each feature addressing a numerical or representational challenge.
        \item The Bayesian extension provides spatially calibrated uncertainty estimates that track model error, enabling principled identification of where the reconstruction should and should not be trusted, potentially enabling us to focus measurements on the areas of larger errors. 
        \item The neural integration formulation achieves time-dependent reconstruction accuracy comparable to static methods, making it a practical tool for modeling evolving galactic potentials from discrete simulation snapshots. This is significant because cosmological simulations are computationally expensive and typically output only a few snapshots across the evolution period. By learning a smooth, continuous representation from these sparse outputs, the framework provides arbitrary temporal resolution between snapshots --- enabling orbit integration, stellar stream modeling, and other dynamical applications that require a smoothly defined potential rather than a discrete set of frozen fields. %
        \item The framework is well-positioned for incorporating into simulation force replay and for direct inference from observational acceleration data. %
        In the former case, it enables differentiable reconstruction of simulated force fields for downstream dynamical studies \citep{Arora_2024, Hunt_2025_SSA}, while in the latter case it can directly ingest pulsar timing and astrometric acceleration measurements without equilibrium assumptions \citep{Chakrabarti+:2021:MeasurementGalacticPlane, Moran+:2024:PulsarbasedMapGalactic, Donlon+:2025:WeighingMilkyWays}. %
    \end{enumerate}
    Together, these results establish physics-informed hybrid modeling as a flexible and physically grounded framework for galactic potential reconstruction, with natural extensions toward observational data and uncertainty-aware dynamical modeling. %


    


\begin{acknowledgments}
    CM was partly supported by the MIT Undergraduate Research Program (UROP).
    NS was supported by The Brinson Foundation through a Brinson Prize Fellowship grant. %
    LN is supported by the Sloan fellowship, and the National Science Foundation under Cooperative Agreement PHY-2019786 (The NSF AI Institute for Artificial Intelligence and Fundamental Interactions, \url{http://iaifi.org/}). %
    The computations in this paper used the Bridges-2 computing cluster at the Pittsburgh Supercomputing Center (PSC). %
    We use the publicly available FIRE-2 cosmological zoom-in simulations \citep{Wetzel2023, wetzel2025secondpublicdatarelease}, from the Feedback In Realistic Environments (FIRE) project, generated using the Gizmo code \citep{Hopkins_2015} and the FIRE-2 physics model \citep{Hopkins2018}.

    This manuscript was prepared with writing assistance from Claude, a large language model (LLM) developed by Anthropic \citep{claude_anthropic}. The LLM was used for lightly editing prose, correcting grammar, and improving clarity of the text. All scientific content and analysis are the sole responsibility of the authors.

\end{acknowledgments}

\begin{contribution}
CM developed and implemented the framework, produced all figures, and wrote 
the manuscript. NS conceived the original research idea, provided scientific 
guidance throughout the project, and edited the manuscript. LN supervised the 
project, provided scientific guidance, and reviewed the manuscript. All authors 
reviewed and approved the final version.

\end{contribution}

\software{
    astropy \citep{2013A&A...558A..33A,2018AJ....156..123A,Astropy+:2022},
    galax, \cite{galax}
    pytreegrav, \cite{pytreegrav}
    unxt \citep{Starkman:2025:UnxtPythonPackage},
    jax, \cite{jax2018github}
    flax, \cite{flax2020github}
    matplotlib, \cite{Hunter:2007}
    numpy, \cite{harris2020array}
    numpyro, \cite{numpyro2019}
    seaborn, \cite{Waskom2021}
    \GalactoPINNS, 
    overcite~\citep{Shariat2026}
}



\appendix

\section{Five-dimensional spherical coordinate system} \label{sec:appendix:5d_coords}

    The five-dimensional (5D) spherical coordinate system of \citet{pinngm} transforms a three-dimensional Cartesian position $\mbf{x} = (x,y,z)^T$ into the five-component input vector $(s,\,t,\,u,\,r_i,\,r_e)$. %
    The transformation is designed to compactify both small and large radii into bounded intervals while preserving smooth directional information, thereby improving the numerical conditioning of the neural network across all spatial scales. %

    The three angular components are the Pines direction cosines \citep{Pines:1973:UniformRepresentation}, %
    \begin{equation}\label{eq:pines_coords}
        s \equiv \frac{x}{r}, \qquad t \equiv \frac{y}{r}, \qquad u \equiv \frac{z}{r},
    \end{equation}
    where $r = \|\mbf{x}\|_2$ is the Euclidean distance from the origin. %
    By construction $s^2 + t^2 + u^2 = 1$, so the triple $(s,t,u)$ encodes the unit direction without the coordinate singularities of the standard spherical representation. %

    A reference radius $r_0$ partitions space into an interior region ($r \leq r_0$) and an exterior region ($r \geq r_0$). For compact bodies, this $r_0$ is typically chosen to be the Brillouin sphere \citep{Brillouin:1933:SphericalHarmonics}, i.e.\ the smallest sphere enclosing the mass distribution. %
    For extended galactic systems, however, no finite enclosing radius exists, so we instead choose $r_0$ to be a characteristic scale of the potential; in the applications considered here, this choice is tied to the spatial non-dimensionalization scale described in \autoref{sec:appendix:scaling}. %
    The two radial coordinates are then defined as %
    \begin{equation}\label{eq:radial_coords}
        r_i \equiv \frac{r}{r_0}, \qquad r_e \equiv \frac{r_0}{r}.
    \end{equation}
    The interior coordinate $r_i \in [0,1]$ maps $r \in [0, r_0]$ to the unit interval, while the exterior coordinate $r_e \in (0,1]$ maps $r \in [r_0,\infty)$ to $(0,1]$. %
    At the reference radius, $r_i = r_e = 1$; outside it $r_e < 1$; inside it $r_i < 1$. %
    Together, $(r_i, r_e)$ span the entire positive real line within a compact, finite domain, eliminating the large dynamic range that would arise from passing a single unscaled $r$ as input. %

    The five inputs are not independent: $s^2 + t^2 + u^2 = 1$ and $r_i r_e = 1$ everywhere.
    Providing all five coordinates as network inputs nonetheless improves conditioning by making both radial regimes equally salient to the optimizer. %
    The transformation is invertible: given $(s,t,u,r_i,r_e)$ and $r_0$, the original position is recovered as $\mbf{x} = r_0\,r_i\,(s,t,u)^T$. %

\section{Input non-dimensionalization} \label{sec:appendix:scaling}

    All inputs are non-dimensionalized with physically consistent scaling factors. %
    To stabilize optimization and ensure that inputs and targets remain physically consistent, we non-dimensionalize positions, accelerations, potentials (and, when applicable, time and velocities) using a single set of derived characteristic scales. %
    We take the spatial scale to be the halo scale radius, $x_\star \equiv r_s$.
    We use this same scale as the reference radius in the 5D spherical coordinate transform, setting $r_0 = x_\star = r_s$. %
    Thus, the transition between the interior and exterior radial coordinates occurs at the characteristic halo scale radius rather than at a finite enclosing radius, which is not well defined for extended galactic potentials. %
    
    We define a characteristic potential scale $u_\star$ from the maximum absolute training potential. %
    When an analytic baseline is included, $u_\star$ is computed from the residual $u_{\rm res} = u - u_{\rm AB}$ so that the scaling reflects the dynamic range the network must learn rather than the dominant analytic component. %
    From $(x_\star,u_\star)$ we derive a consistent dynamical time scale, $t_\star \equiv \sqrt{x_\star^2/u_\star}$, which follows from dimensional analysis of $\phi \sim x^2/t^2$ in gravitational units. %
    This in turn fixes the acceleration and velocity scales as $a_\star \equiv x_\star/t_\star^2$ and $v_\star \equiv x_\star/t_\star$. %
    We then apply uniform affine scalers to map each quantity to a bounded range (here $[-1,1]$), using the same $(x_\star,t_\star,u_\star)$ to scale all channels. %
    In the time-dependent setting, the time coordinate is scaled by $t_\star$ and concatenated to the scaled spatial coordinates, ensuring that the learned evolution operates in dimensionless units compatible with the scaled potential and acceleration targets. %

\section{Training details} \label{sec:appendix:training_details}

\subsection{Training schedules}
    For all training schedules, we use the Adam optimizer with an exponential decay in the learning rate. We find that an initial learning rate of 0.003 with a learning-rate decay factor of 0.5 every $2{,}000$ epochs generally yields the best performance. %

    When trainable analytic parameters are included (PINN~\ref{model:pinnV}), training proceeds in 
    two stages. An initial analytic-focused phase fits the baseline parameters to capture 
    the dominant large-scale structure of the potential, followed by a joint optimization phase 
    in which both the analytic parameters and neural residual are refined together. We found that reserving the first $\sim 20\%$ of the total training epochs for focused analytic inference generally yields the best results. In practice, the appropriate training schedule depends on the complexity of the analytic baseline, the number of free parameters, and whether accurate parameter 
    recovery or overall reconstruction accuracy is the primary goal.
    
    Under Bayesian inference, the same two-stage strategy is applied to the variational 
    posterior. In stage~1, we use a broad prior on the analytic parameters ($\sigma_\alpha$ = 0.5) and a tight prior on the residual network weights
    ($\sigma_\theta = \sigma_{\text{det}}/10$), where $\sigma_{\text{det}}$ 
    is the root mean square of the weights of the pre-trained deterministic network. This stage encourages the analytic baseline to absorb 
    large-scale structure before the residual field is free to vary. In stage~2, a tightened weight prior and relaxed residual prior ($\sigma_\theta = \sigma_{\text{det}}$) and tighter prior on the analytic parameters ($\sigma_\alpha$ = 0.001) allow the network to 
    capture higher-order corrections while the analytic parameters remain well-constrained. 
    Each stage is initialized from the previous variational solution to ensure continuity in the posterior evolution.  %
    Specific epoch counts and prior values for each experiment are given in the corresponding results sections. %

\subsection{Importance weighting \label{sec:appendix:importance}}
    For PINN models that include a fixed analytic baseline  $\PhiAB(\mbf{x})$, it can be valuable to focus training capacity on regions where this baseline is least accurate. To this end, we optionally employ a per-point importance weighting scheme. For each training position $\mathbf{x}_i$, we compute a scalar weight $w_i \equiv \|\mathbf{a}(\mathbf{x}_i)-\mathbf{a}_{\rm AB}(\mathbf{x}_i)\|$, where $\mathbf{a}_{\rm AB}(\mathbf{x}_i)\equiv -\nabla \PhiAB(\mathbf{x}_i)$ is the acceleration predicted by the analytic baseline. The acceleration training objective in \autoref{eq:model_design:architecture:loss} is then replaced by the weighted mean
    $$\mathcal{L}_{\text{acc}}(\boldsymbol{\theta}) = \frac{1}{N}\sum_{i=1}^{N} w_i  \Bigg(
                \left\| -\nabla \phi(\mathbf{x}_i \mid \boldsymbol{\theta}) - \mathbf{a}_i \right\| \\
                + \lambda_{\mathrm{r}}
                \frac{
                    \left\| -\nabla \phi(\mathbf{x}_i \mid \boldsymbol{\theta}) - \mathbf{a}_i \right\|
                }{
                    \left\| \mathbf{a}_i \right\| + \epsilon_{\mathrm{mach}}
                }
            \Bigg) \; .$$
    
    Points where the analytic model is a poor approximation receive larger weights, biasing the network to correct precisely where a correction is most needed. 

\subsection{Additional implementation details}
    We use \texttt{gelu} activation between layers, which empirically yields better performance than \texttt{softplus} and \texttt{relu} activation. This finding aligns with other physics-informed frameworks which report improved accuracy using \texttt{gelu} \citep{wang2023expertsguidetrainingphysicsinformed}. %
    
    All training was performed on an Apple M1 CPU; the \GalactoPINNS code is fully GPU-compatible, which would reduce training times by approximately an order of magnitude for most systems.

    The training datasets, evaluation datasets, and saved model parameters used in this work are available on Zenodo at \doi{10.5281/zenodo.20361367}.

\section{Evaluation details} \label{sec:appendix:eval_details}
    We evaluate the performance using two primary metrics, which are designed to probe both the local force accuracy and global dynamical consistency. The metrics are defined as:
    \begin{itemize}[leftmargin=*]
        \item Mean acceleration error (MAE) (reported in \%):
            \begin{equation}
                \mathrm{MAE}\, \;=\;
                100 \times \frac{1}{N}\sum_{i=1}^{N}
                \frac{\left\lVert \mathbf{a}_{\text{pred}}(\mathbf{x}_i) - \mathbf{a}_{\text{true}}(\mathbf{x}_i) \right\rVert}
                {\left\lVert \mathbf{a}_{\text{true}}(\mathbf{x}_i) \right\rVert}
            \end{equation}
        where $N$ is the number of evaluation points, $\mathbf{a}_{\text{pred}}(\mathbf{x}_i)$ is the predicted acceleration at position $\mathbf{x}_i$, and $\mathbf{a}_{\text{true}}(\mathbf{x}_i)$ is the true acceleration at the same position.
        \item Mean time-averaged orbit deviation  (MOD) (reported in $\mathrm{kpc}$):
            \begin{equation}
                \mathrm{MOD} \;=\;
                \frac{1}{D}\sum_{d=1}^{D}\;
                \frac{1}{|\mathcal{T}|}\sum_{t \in \mathcal{T}}
                \left\lVert \mathbf{r}_{d, \text{pred}}(t) - \mathbf{r}_{d, \text{true}}(t) \right\rVert\,
            \end{equation}
        where $D$ is the number of evaluation orbits, $\mathcal{T}$ is the set of saved times in each of these orbits, $\mathbf{r}_{d, \text{pred}}(t)$ is the predicted location of the orbit at time $t \in \mathcal{T}$, and $\mathbf{r}_{d, \text{true}}(t)$ is the true location of the orbit at the same time.
    \end{itemize}

    Orbit fidelity (measured by the MOD metric) provides a complementary probe of force reconstruction accuracy (measured by the MAE metric). %
    While orbit-based and acceleration-based metrics are not independent, they assess consistency at different scales: pointwise acceleration errors quantify local, small-scale accuracy, whereas orbit deviations probe the global structure of the potential and its long-term dynamical consistency. %

\begin{sidewaysfigure*}
    \vspace*{1cm} 
    \hspace*{-1.0cm} 
    \begin{minipage}{0.54\textheight}
        \centering
        \includegraphics[width=\linewidth]{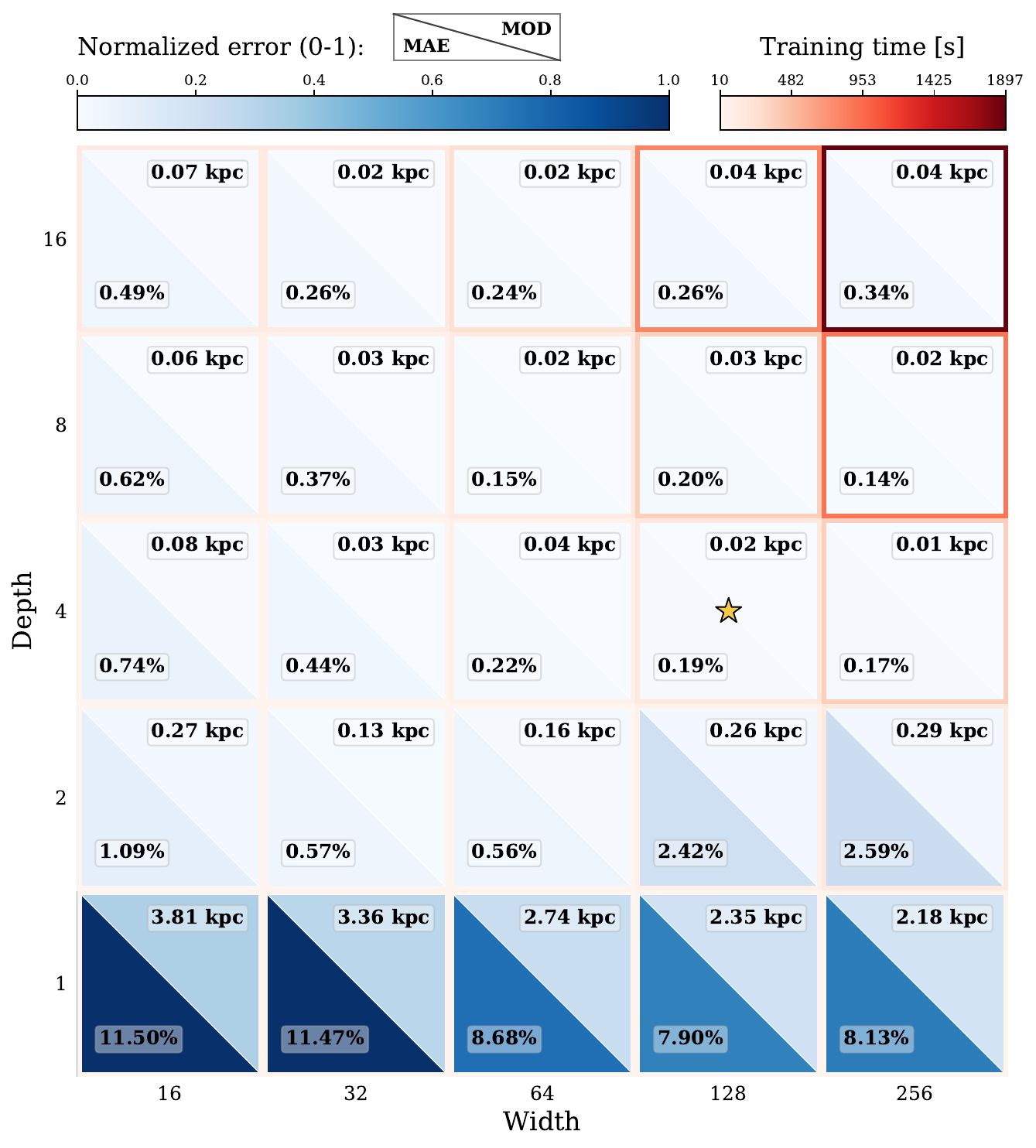}
    \end{minipage}
    \hfill
    \begin{minipage}{0.56\textheight}
        \centering
        \includegraphics[width=\linewidth]{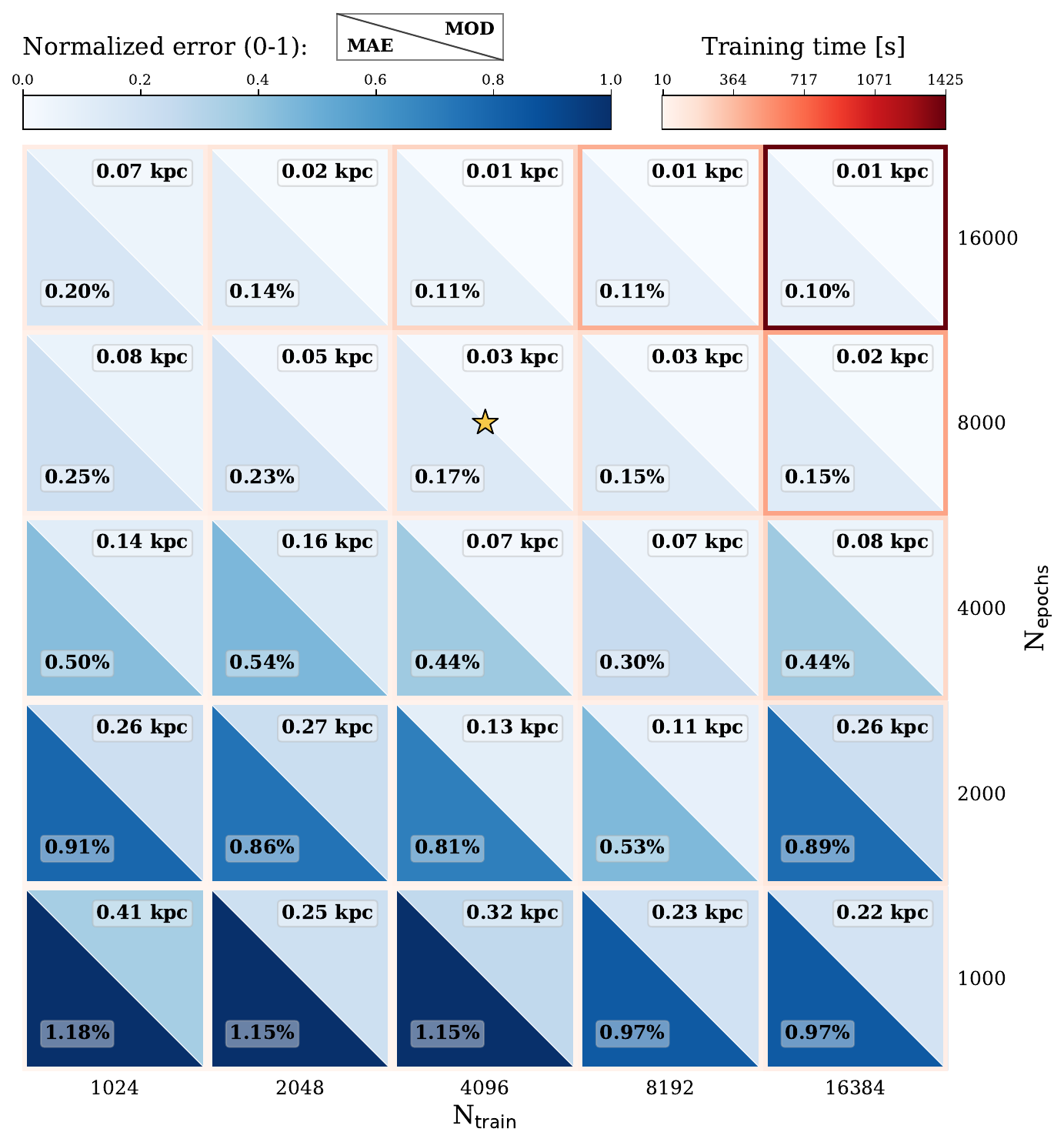}
    \end{minipage}
    
    \caption{\textbf{Architecture and training ablations.} %
    Each cell shows the performance of one training configuration, with the diagonal split separating two diagnostics: the lower-left triangle gives the mean acceleration error (MAE; $\%$), evaluated on 32,768 test points, while the upper-right triangle gives the mean orbit deviation (MOD; kpc), computed from 150 test-particle orbits initialized at randomly sampled positions and integrated in the learned potential for $500\,\mathrm{Myr}$. %
    The color of each cell border encodes the training time, and the star marks the configuration selected for follow-up experiments. %
    \textit{Left: Network depth--width ablation.} %
    PINN~\ref{model:pinnV} models are trained with different network depths and widths on the \gls{MW}--\gls{LMC} test system. %
    \textit{Right: Training-set size and epoch-count ablation.} %
    PINN~\ref{model:pinnIV} models with fixed architecture are trained with different numbers of training points, $N_\mathrm{train}$, and training epochs, $N_\mathrm{epochs}$. %
    Performance is evaluated on the same test set as in the left panel. %
    }
    \label{fig:ablation_summary}
    \end{sidewaysfigure*}

    In the left panel of \autoref{fig:ablation_summary}, we report the performance of a PINN~\ref{model:pinnV} model with various combinations of network depth and width. As the benchmark test system, we use the \gls{MW}--\gls{LMC} dataset described in \autoref{sec:results:static_mw_lmc}. All models are trained with the same training set of $4{,}096$ points, and the same training schedule ($2{,}000$ analytic-focused epochs, followed by $10{,}000$ residual-focused epochs).
    Each grid cell in \autoref{fig:ablation_summary} represents one network configuration, in which we display the two performance metrics: MAE and MOD, separated by the diagonal. We also represent the training time for each model in the color of the cell border. The performance in both metrics generally improves moving towards the upper right of the figure (larger depth and width), and the color of the border becomes darker (corresponding to longer training times). This trend highlights the trade-off between increasing accuracy and increasing training time. However, the most complex model (upper right) performs slightly worse in both metrics than the two adjacent models. This is explained by the fact that all models are trained for the same number of epochs ($12{,}000$) rather than training to convergence. For the most complex model, this training length may not be sufficient to reach the highest accuracy, which explains the slight worsening relative to shallower models. 
    Notice that the two metrics (MAE and MOD) track each other closely, as expected: 
    accurate force reconstruction is a necessary condition for accurate orbit integration, 
    so improvements in pointwise acceleration errors translate into better long-term orbital fidelity. 
    Performance also depends much more strongly on depth than width, with widths of 64 already sufficient to saturate the accuracy for any given depth. For practical purposes, we find that a depth of 4 and width of 128 achieves high accuracy (with mean acceleration error $< 0.2 \%$) and fast training ($\sim 5$ minutes on an Apple M1 CPU); this combination is marked with a star in the left panel of \autoref{fig:ablation_summary}. 

    In the right panel of \autoref{fig:ablation_summary}, we perform a similar comparison, but instead of varying the size of the network, we vary the number of training epochs and the number of points in the training set. Here, all models have fixed network size, corresponding to the starred configuration from the depth-width ablation: depth 4 and width 128. To simplify the training schedule, we train a PINN~\ref{model:pinnIV} model using a fixed analytic baseline, rather than PINN V. 
    Performance generally improves moving towards the upper right corner of the figure: with sufficient training points to constrain the solution, accuracy improves with longer training schedules. For fewer total epochs ($2{,}000$ and $4{,}000$), performance is slightly degraded by increasing the number of training points. This suggests that larger training sets require longer optimization to be effective; with too few epochs, the model may underfit the expanded set of training points. For this particular system, we find that $8{,}000$ epochs and $4{,}096$ training points provides a good balance of accuracy ($< 0.2 \%$ MAE) and efficiency (training time $\sim$ 5 minutes). This combination is marked with a star on the right panel of \autoref{fig:ablation_summary}. Note that this is not a general prescription for the optimal configuration; for more complex systems, more training points and longer training schedules may be required.
        

    \clearpage
    \clearpage

\bibliographystyle{aasjournal}   
\bibliography{main}   

\end{document}

%% file: figure_schematic.tex
\begin{figure*}
\centering
\resizebox{\textwidth}{!}{%
\begin{tikzpicture}
    \useasboundingbox (-5,-6) rectangle (14.5,2);

    \node[inner sep=0] (A) at (-3,-1.5)
        {\includegraphics[width=5.0cm]{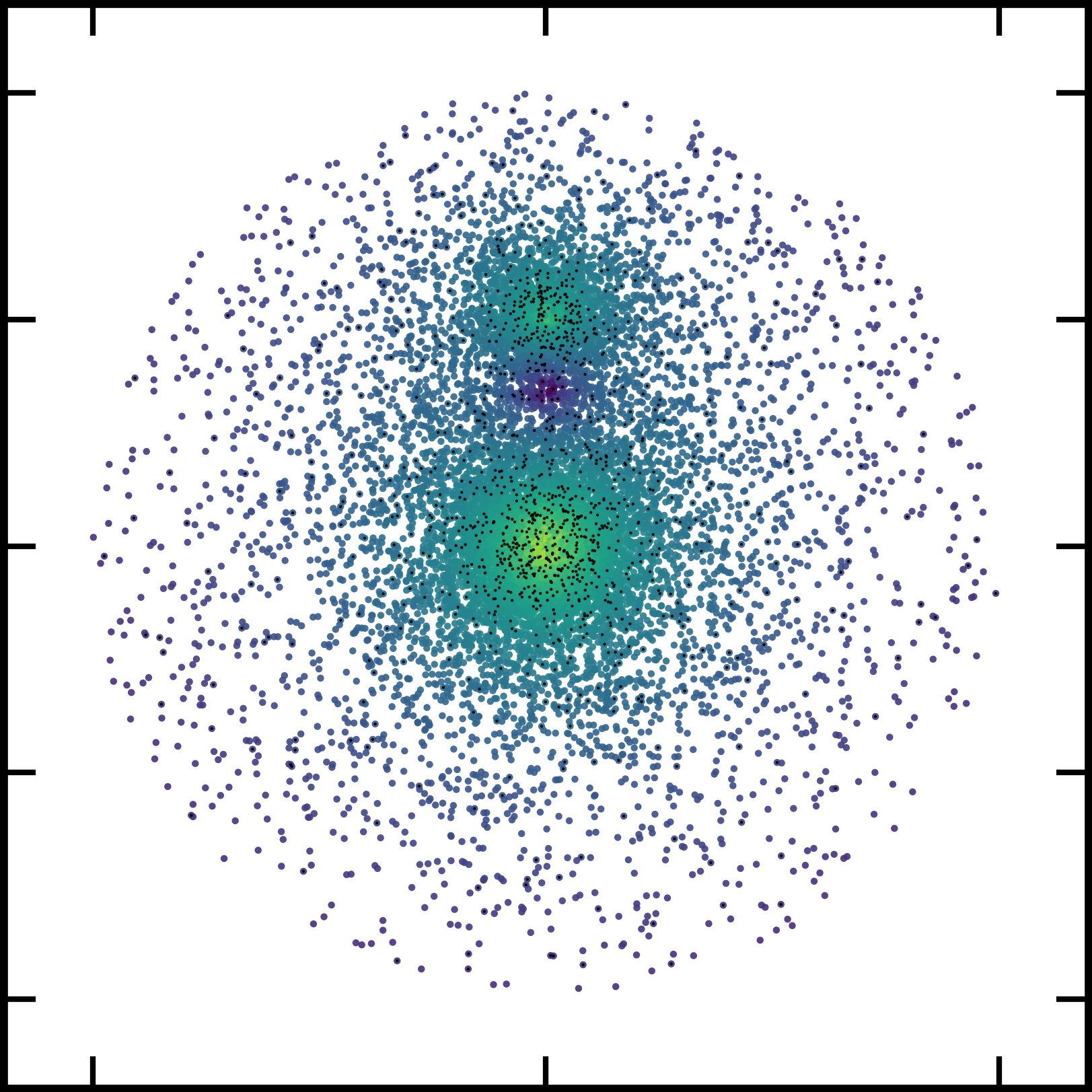}};

    \node[above=4pt of A.north, font=\small\bfseries]
        {\shortstack[c]{3D snapshot of\\ evolving potential}};

    \draw[red, thick]
        ($(A.center)+(-0.5cm,-0.1cm)$) rectangle
        ($(A.center)+(0.5cm,0.1cm)$);

    \coordinate (roiTL) at ($(A.center)+(-0.5cm, 0.1cm)$);
    \coordinate (roiTR) at ($(A.center)+( 0.5cm, 0.1cm)$);
    \coordinate (roiBL) at ($(A.center)+(-0.5cm,-0.1cm)$);
    \coordinate (roiBR) at ($(A.center)+( 0.5cm,-0.1cm)$);

    \newsavebox{\Azbox}
    \newlength{\AzW}
    \newlength{\AzH}
    \setlength{\AzW}{4.5cm}
    \sbox{\Azbox}{\includegraphics[width=\AzW]{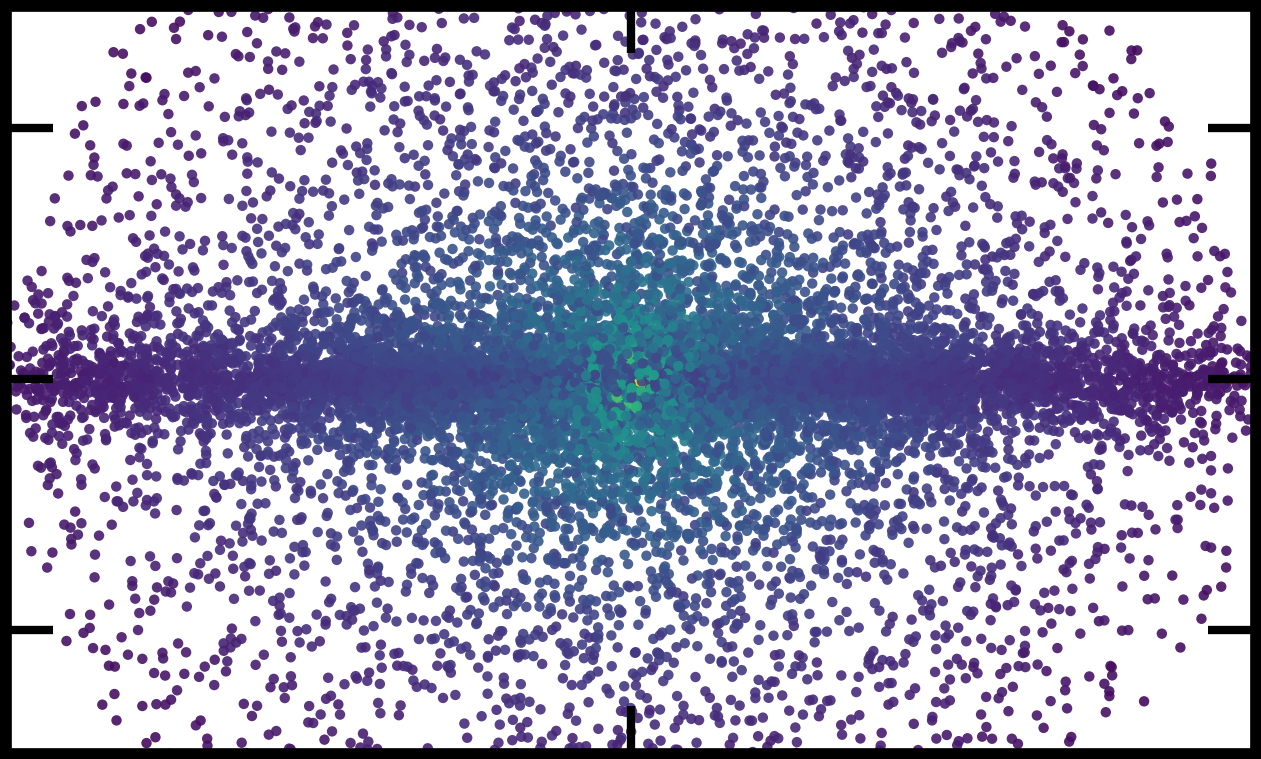}}
    \setlength{\AzH}{\ht\Azbox}

    \node[inner sep=0] (Az) at (-3.5,-4.1) {\usebox{\Azbox}};

    \coordinate (AzTL) at ($(Az.center)+(-0.5*\AzW,  0.5*\AzH)$);
    \coordinate (AzTR) at ($(Az.center)+( 0.5*\AzW,  0.5*\AzH)$);
    \coordinate (AzBL) at ($(Az.center)+(-0.5*\AzW, -0.5*\AzH)$);
    \coordinate (AzBR) at ($(Az.center)+( 0.5*\AzW, -0.5*\AzH)$);

    \draw[red, thick] (roiBL) -- (AzTL);
    \draw[red, thick] (roiTR) -- (AzTR);

    \node[
        box,
        fill=gray!30,
        font=\small,
        text width=3.7cm,
        anchor=west,
        inner sep=2pt,
        rounded corners=2pt,
        align=left
    ] (B) at (0.7,-1.9) {%
        \begin{tabular}{@{}l@{}c@{\hspace{2pt}}l@{}}
            n & : & scaling function \\
            $\phi_{\mathrm{AB}}$ & : & analytic baseline
        \end{tabular}
    };
    \node[above=2pt of B.north, font=\small\bfseries] {\textbf{Physics-informed priors}};

    \path let \p1 = (A.north) in
        node[
            box, fill=gray!30, font=\small, text width=3.7cm,
            anchor=north west, inner sep=2pt, rounded corners=2pt
        ] (C) at (0.7,\y1 -3mm) {};

    \draw ($(C.west)+(0.4cm,0.30cm)$) -- ($(C.west)+(0.4cm,-0.30cm)$);
    \node[anchor=west, font=\small] at ($(C.west)+(2pt,0)$) {$t$};
    \node[anchor=west, font=\small] at ($(C.west)+(0.4cm+4.2pt,0)$) {5D spherical coords.};

    \node[bigboxfixed, font=\small, anchor=west, inner sep=6pt] (F) at (6.1,0.4) {};

    \node[bigboxfixed, font=\small, anchor=west, inner sep=6pt] (D) at (6.1,-1.8) {%
        \(
        \begin{aligned}
            \phi(\mathbf{x}, t)
            &= \frac{1}{n(\mathbf{x}, t)} \cdot
            \Bigl[
            \tilde{\phi}_{\mathrm{NN}_0}(\mathbf{x} \!\mid\! \boldsymbol{\theta}_1)
            + \int_{0}^{t}\! \dot{\tilde{\phi}}_\mathrm{NN}(\mathbf{x}, t' \!\mid\! \boldsymbol{\theta}_2)\,
                \mathrm{d}t'
            \Bigr] \\
            &\quad + \phi_{\mathrm{AB}}(\mathbf{x}, t\mid\boldsymbol{\theta}_3)
        \end{aligned}
        \)
    };

    \node[bigboxfixed, font=\small, anchor=west, inner sep=8pt] (E) at (6.1,-4.4) {%
        \begin{equation*}
            \mathcal{L}(\boldsymbol{\theta})
            = \frac{1}{N} \sum_{i=1}^{N}
            \Biggl(
                \left\| -\nabla \phi(\mathbf{x}_i \mid \boldsymbol{\theta}) - \mathbf{a}_i \right\| 
                + \lambda_{\mathrm r}
                \frac{
                    \left\| -\nabla \phi(\mathbf{x}_i \mid \boldsymbol{\theta}) - \mathbf{a}_i \right\|
                }{
                    \left\| \mathbf{a}_i \right\|
                }
            \Biggr)
        \end{equation*}
    };

    \node[anchor=north, font=\small\bfseries] at ($(E.east)+(-45pt,40pt)$) {Autodiff. +  Loss};
    \node[anchor=west, font=\small\bfseries] at ($(D.east)+(-70pt,34pt)$) {Reconstruct $\phi$};
    \node[anchor=north, font=\small\bfseries] at ($(F.east)+(-18pt,34pt)$) {Sample};

    \path let \p1 = (C.west), \p2 = (A.east) in
        coordinate (Aproj) at (\x2,\y1);
    \draw[opacity=0] (A.east) -- (Aproj);
    \draw[boldarrow] (Aproj) -- (C.west)
        node[midway, above=4pt] {$\mathbf{x}(t)$};

    \draw[boldarrow] (B) -- (B -| F.west);
    \draw[boldarrow] (C) -- (C -| D.west);

    \draw[boldarrow] (F) -- (D);
    \draw[boldarrow] (D) -- (E);

    \draw[boldarrow]
        ($(A.south)+(56pt,0)$) |- (E)
        node[pos=0.75, above=4pt] {$\mathbf{a}(t)$};

    \begin{scope}[xshift=1.2cm, yshift=-0.4cm]
        \coordinate (w1L) at (6.2,0.5);
        \coordinate (w1P) at (7.0,1.1);
        \coordinate (w1R) at (7.4,0.5);
        \draw[thick] (w1L) to[out=0,in=180] (w1P) to[out=0,in=180] (w1R);
        \coordinate (w1mid) at ($(w1L)!0.5!(w1R)$);
        \node at ($(w1mid)+(2pt,5pt)$) {$\boldsymbol{\theta}_1$};

        \coordinate (w2L) at (8.7,0.5);
        \coordinate (w2P) at (9.3,1.1);
        \coordinate (w2R) at (9.9,0.5);
        \draw[thick] (w2L) to[out=0,in=180] (w2P) to[out=0,in=180] (w2R);
        \coordinate (w2mid) at ($(w2L)!0.5!(w2R)$);
        \node at ($(w2mid)+(0,5pt)$) {$\boldsymbol{\theta}_2$};

        \coordinate (w3L) at (11.2,0.5);
        \coordinate (w3H1) at (11.6,1.0);
        \coordinate (w3Low) at (11.7,0.9);
        \coordinate (w3H2) at (12.0,1.2);
        \coordinate (w3R) at (12.4,0.5);
        \draw[thick] (w3L) to[out=0,in=180] (w3H1) to[out=0,in=180] (w3Low)
                           to[out=0,in=180] (w3H2)  to[out=0,in=180] (w3R);
        \coordinate (w3mid) at ($(w3L)!0.5!(w3R)$);
        \node at ($(w3mid)+(0,5pt)$) {$\boldsymbol{\theta}_3$};
    \end{scope}
\end{tikzpicture}
}

\caption{%
    \textbf{Model schematic}, highlighting the core elements of the method. %
    The framework takes as input particle positions $\mathbf{x}(t)$ and gravitational accelerations $\mathbf{a}(t)$ sampled from the target potential. %
    Positions and time are passed through a coordinate transformation and enter the reconstruction through three learned components parameterized by $\boldsymbol{\theta}_1$, $\boldsymbol{\theta}_2$, and $\boldsymbol{\theta}_3$, shown here as posterior distributions under Bayesian inference.
    These components parameterize the initial spatial correction, the learned time derivative, and the analytic baseline, respectively, and are combined with the physics-informed priors -- spatial scaling $n(\mathbf{x}, t)$ and analytic baseline $\phi_\mathrm{AB}(\mathbf{x}, t)$ -- to reconstruct the full time-dependent potential $\phi(\mathbf{x}, t)$. %
    The reconstructed potential is differentiated and compared to the true accelerations $\mathbf{a}(t)$ through the loss function. %
}
\label{fig:schematic}
\end{figure*}